\documentclass{llncs}
\usepackage{makeidx}  
\usepackage{color}
\usepackage{graphicx}
\usepackage{float}

\usepackage{amsmath}
\usepackage{amssymb}
\usepackage[labelformat=simple]{subfig}
\usepackage[ruled,lined,linesnumbered]{algorithm2e}
\usepackage{algorithmic}
\usepackage{blindtext}
\usepackage{enumitem}
\usepackage{multirow}
\usepackage[labelformat=simple]{subfig}
\usepackage[pagebackref=true,breaklinks=true,letterpaper=true,colorlinks,bookmarks=false]{hyperref}
\usepackage{threeparttable/threeparttable}
\usepackage{bm}
\usepackage{romannum}
\usepackage[normalem]{ulem}
\usepackage[font={footnotesize}]{caption}
\usepackage{mathabx}
\usepackage[table]{xcolor}
\usepackage{multirow, makecell}
\usepackage{xcolor}
\usepackage{colortbl}
\usepackage{multibib}
\usepackage[misc,geometry]{ifsym}

\definecolor{maroon}{cmyk}{0,0.87,0.68,0.32}
\definecolor{darkgreen}{rgb}{0,0.55,0}
\usepackage{amssymb}
\usepackage{pifont}
\newcommand{\cmark}{\ding{51}}%
\newcommand{\xmark}{\ding{55}}%
\newcommand{\etal}{\mbox{\emph{et al.\ }}}
\newcommand{\ie}{\mbox{\emph{i.e.,\ }}}
\newcommand{\eg}{\mbox{\emph{e.g.,\ }}}
\definecolor{mygray}{gray}{0.3}
\newcommand{\tabincell}[2]{\begin{tabular}{@{}#1@{}}#2\end{tabular}}

\usepackage{array}
\newcolumntype{P}[1]{>{\centering\arraybackslash}p{#1}}

\usepackage{soul}

\newenvironment{jlblue}{\color{black}}{}
\newenvironment{jlmagenta}{\color{black}}{}

\definecolor{br}{rgb}{0.5, 0.2, 0.0}
\newenvironment{ntbrown}{\color{br}}{}

\makeatletter
\newcommand\footnoteref[1]{\protected@xdef\@thefnmark{\ref{#1}}\@footnotemark}
\makeatother

\newenvironment{zzred}{\color{black}}{}

\definecolor{dr}{rgb}{0.6, 0.0, 0.1}
\newenvironment{rbfcolor}{\color{black}}{}

\newcommand{\repcite}[1]{}
\newcommand{\repciteapp}[1]{}

\begin{document}
\pagestyle{plain}
\thispagestyle{plain}

\mainmatter              
\title{Models Genesis: Generic Autodidactic Models for 3D Medical Image Analysis}

\author{
Zongwei Zhou\inst{1} \and
Vatsal Sodha\inst{1} \and
Md Mahfuzur Rahman Siddiquee\inst{1} \and \\
Ruibin Feng\inst{1} \and
Nima Tajbakhsh\inst{1} \and
Michael B. Gotway\inst{2} \and
Jianming Liang\inst{1}
}
%
\authorrunning{Z. Zhou et al.}
%
\institute{
Arizona State University, Scottsdale, AZ 85259 USA \\
\email{\{zongweiz,vasodha,mrahmans,rfeng12,ntajbakh,jianming.liang\}@asu.edu} \and 
Mayo Clinic, Scottsdale, AZ 85259 USA \\
\email{Gotway.Michael@mayo.edu}
}

\maketitle              
\setcounter{footnote}{0}

\begin{abstract}
    Transfer learning from {\em natural} image to {\em medical} image has established as one of the most practical paradigms in deep learning for medical image analysis. However, to fit this paradigm, 3D imaging tasks in the most prominent imaging modalities ({\em e.g.}, CT and MRI) have to be reformulated and solved in 2D, losing rich 3D anatomical information and inevitably compromising the performance. To overcome this limitation, we have built a set of models, called Generic Autodidactic Models, nicknamed Models Genesis, because they are created {\em ex nihilo} (with no manual labeling), self-taught ({\zzred learned} by self-supervision), and generic (served as source models for generating application-specific target models). Our extensive experiments demonstrate that our Models Genesis significantly outperform learning from scratch in all five target 3D applications covering both segmentation and classification.
    More importantly, learning a model from scratch simply in 3D may not necessarily yield performance better than transfer learning from ImageNet in 2D, but our Models Genesis consistently top any 2D approaches including fine-tuning the models pre-trained from ImageNet as well as fine-tuning the 2D versions of our Models Genesis, confirming the importance of 3D anatomical information and significance of our Models Genesis for 3D medical imaging. This performance is attributed to our unified self-supervised learning framework, built on a simple yet powerful observation: the sophisticated yet recurrent anatomy in medical images can serve as strong supervision signals for deep models to learn common anatomical representation automatically via self-supervision. 
    As open science, all pre-trained Models Genesis are available at \href{https://github.com/MrGiovanni/ModelsGenesis}{https://github.com/MrGiovanni/ModelsGenesis}.
\end{abstract}

\begin{table}[t]
\scriptsize
\centering
\begin{threeparttable}
\caption{Target tasks.}
\label{tab:terminology}
    \begin{tabular}{p{0.06\linewidth}p{0.22\linewidth}p{0.11\linewidth}p{0.115\linewidth}p{0.45\linewidth}}
        \hline
        Code$^{\dagger}$ & Object & Modality & Source  & Description \\
        \hline
        \texttt{NCC} & Lung Nodule & CT & \tiny{\href{https://luna16.grand-challenge.org/}{LUNA2016}} & \scriptsize{Lung nodule false positive reduction} \\
        \texttt{NCS} & Lung Nodule & CT & \tiny{\href{https://wiki.cancerimagingarchive.net/display/Public/LIDC-IDRI}{LIDC-IDRI}} & \scriptsize{Lung nodule segmentation} \\
        \texttt{ECC} & Pulmonary Embolism & CT & \tiny{\href{https://link.springer.com/chapter/10.1007/978-3-319-24571-3_8}{PE-CAD}}  & \scriptsize{Pulmonary embolism false positive reduction} \\
        \texttt{LCS} & Liver & CT & \tiny{\href{https://competitions.codalab.org/competitions/17094}{LiTS2017}} & \scriptsize{Liver segmentation} \\
        \texttt{DXC} & Pulmonary Diseases & X-ray & \tiny{\href{https://nihcc.app.box.com/v/ChestXray-NIHCC}{ChestX-ray8}} & {\scriptsize Eight pulmonary diseases classification} \\ 
        \texttt{IUC} & CIMT RoI & Ultrasound & \tiny{\href{https://www.ncbi.nlm.nih.gov/pubmed/20381661}{UFL MCAEL}} & \scriptsize{RoI, bulb, and background classification} \\
        \texttt{BMS} & Brain Tumor & MRI & \tiny{\href{https://www.smir.ch/BRATS/Start2013}{BraTS2013}} & \scriptsize{Brain tumor segmentation} \\
        \hline
    \end{tabular}
    \begin{tablenotes}
        \scriptsize
        \item $^{\dagger}$ The first letter denotes the object of interest (``\texttt{N}'' for lung nodule, ``\texttt{E}'' for pulmonary embolism, ``\texttt{L}'' for liver, etc); the second letter denotes the modality (``\texttt{C}'' for CT, ``\texttt{X}'' for X-ray, ``\texttt{U}'' for Ultrasound, etc);  the last letter denotes the task (``\texttt{C}'' for classification, ``\texttt{S}'' for segmentation).
    \end{tablenotes}
\end{threeparttable}
\end{table}

\section{Introduction}
\label{sec:introduction}

Given the marked differences between {\em natural} images and {\em medical} images, we hypothesize that transfer learning can yield more powerful (application-specific) {\em target} models if the {\em source} models are built directly from medical images. To test this hypothesis, we have chosen chest imaging because the chest contains several  critical organs, which are prone to a number of diseases that result in substantial morbidity and mortality and thus are associated with significant health-care costs. 
In this research, we focus on Chest CT, because of its prominent role in diagnosing lung diseases, and our research community has accumulated several Chest CT image databases, for instance, LIDC-IDRI\footnote{\label{foot:LIDC}\href{https://wiki.cancerimagingarchive.net/display/Public/LIDC-IDRI}{https://wiki.cancerimagingarchive.net/display/Public/LIDC-IDRI}} and NLST\footnote{\href{https://biometry.nci.nih.gov/cdas/nlst/}{https://biometry.nci.nih.gov/cdas/nlst/}}, containing a large number of Chest CT images.
Therefore, we seek to answer the following question: {\em Can we utilize the large number of available Chest CT images without systematic annotation to train source models that can yield high-performance target models via transfer learning?} 

To answer this question, we have developed a framework that trains generic, source models for 3D  imaging. 
We call the models trained with our framework  Generic Autodidactic Models, nicknamed Models Genesis, and refer to the model trained using Chest CT scans as Genesis Chest CT. 
As ablation studies, we have also trained a downgraded 2D version using 2D Chest CT slices, called Genesis Chest CT 2D. 
To demonstrate the effectiveness of Models Genesis in 2D applications, we have trained a 2D model based on ChestX-ray8\footnote{\label{foot:xray8}\href{https://nihcc.app.box.com/v/ChestXray-NIHCC}{https://nihcc.app.box.com/v/ChestXray-NIHCC}}, named as Genesis Chest X-ray.

Our extensive experiments detailed in Sec.~\ref{sec:experiments} demonstrate that Models Genesis, including Genesis Chest CT, Genesis Chest CT 2D, and Genesis Chest X-ray, {\em significantly} outperform learning from scratch in all seven target tasks (see \tablename~\ref{tab:terminology}). As revealed in \tablename~\ref{tab:2D_3D_target_tasks}, learning from scratch simply in 3D may {\em not} necessarily yield performance better than fine-tuning state-of-the-art ImageNet models, but our Genesis Chest CT {\em consistently} top any 2D approaches including fine-tuning ImageNet models as well as fine-tuning our Genesis Chest X-ray and Genesis Chest CT 2D, confirming the importance of 3D anatomical information in Chest CT and significance of our self-supervised learning method in 3D medical image analysis.

\begin{figure}[t]
\centering
    \includegraphics[width=0.97\linewidth]{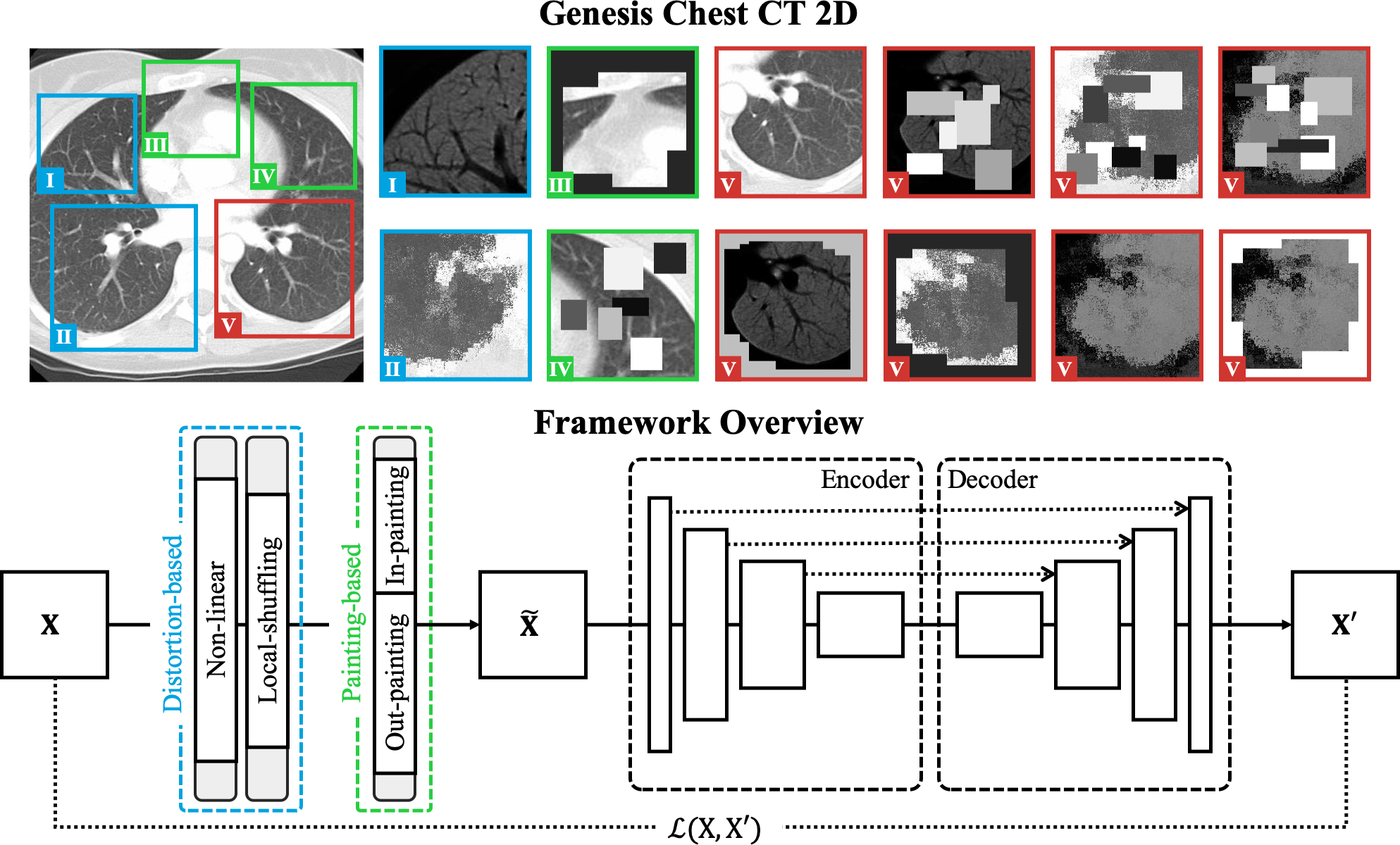}
    \caption{
    Our unified self-supervised learning framework consolidates four novel transformations: 
    I) non-linear, 
    II) local-shuffling, 
    III) out-painting, and 
    IV) in-painting into a single image restoration task. 
    Specifically, each arbitrarily-size patch $\text{X}$ cropped at random location from an unlabeled image can undergo at most three of above transformations, resulting in a transformed patch $\tilde{\text{X}}$ (see I--V). Note that out-painting and in-painting are mutually exclusive.
    For simplicity and clarity, we illustrate our idea on a 2D CT slice, but our Genesis Chest CT is trained using 3D images directly.
    A Model Genesis, an encoder-decoder architecture, is trained to learn a common visual representation by restoring the original patch $\text{X}$ (as ground truth) from the transformed one $\tilde{\text{X}}$ (as input), aiming to yield high-performance target models. 
    }
\label{fig:unified_framework}
\end{figure}

This performance is attributable to the following key observation: 
medical imaging protocols typically focus on particular parts of the body for specific clinical purposes, resulting in images of similar anatomy. The sophisticated yet recurrent anatomy offers consistent patterns for self-supervised learning to discover common  representation of a particular body part (the lungs in our case). The fundamental idea behind our unified self-supervised learning method as illustrated in \figurename~\ref{fig:unified_framework} is to recover anatomical patterns from images transformed via various ways in a unified framework.

\section{Models Genesis}

\jlmagenta Models Genesis learn from scratch on unlabeled images, with an objective to yield a common visual representation that is generalizable and transferable across diseases, organs, and modalities. In Models Genesis, an encoder-decoder, as shown in \figurename~\ref{fig:unified_framework}, is trained using a series of self-supervised schemes. Once trained, the encoder alone can be fine-tuned for target classification tasks; while the encoder and decoder together can be for target segmentation tasks. 
For clarity, we formally define a {\em training scheme} as the process that transforms patches with any of the  transformations, as illustrated in \figurename~\ref{fig:unified_framework}, and trains a model to restore the original patches from the transformed counterparts.
In the following, we first explain each of our self-supervised learning schemes with its learning objectives and perspectives, followed by a summary of the four unique properties of our Models Genesis. Along the way, we also contrast Models Genesis with existing approaches to show our {\bf innovations} and {\bf novelties}.

\begin{itemize}
    \item[$\bullet$] {\bf Learning  appearance via non-linear transformation.}
    Absolute or relative intensity values in  medical images convey important information about the imaged structures and organs. For instance,
    the Hounsfield Units in CT scans correspond to specific substances of the human body. As such, intensity information can be used as a strong source of pixel-wise supervision. 
    To preserve relative intensity information of anatomies during image transformation, we use B{\'e}zier Curve, a smooth and monotonous transformation function, which assigns every pixel a unique value, ensuring a one-to-one mapping. Restoring image patches distorted with non-linear transformation focuses Models Genesis on learning organ appearance (shape and intensity distribution).
    \figurename~\ref{fig:unified_framework}--\Romannum{1} shows examples of the transformed images. Due to limited space, we provide the implementation details in Appendix\footnoteref{foot:Appendix}~Sec.~\ref{sec:nonlinear_appendix}. 
    
    \smallskip
    \item[$\bullet$] {\bf Learning texture via local pixel shuffling.}
    Given an original patch, local pixel shuffling consists of sampling a random window from the patch followed by shuffling the order of contained pixels resulting in a transformed patch.
    The size of the local window determines the task difficulty, but we keep it smaller than the model's receptive field, and also small enough to prevent changing the global content of the image. Note that our method is quite different from PatchShuffling~\cite{kang2017patchshuffle}, which is a regularization technique to avoid over-fitting. {\jlmagenta To recover from local pixel shuffling,  Models Genesis must memorize local boundaries and texture.}
    Examples of local-shuffling are illustrated in~\figurename~\ref{fig:unified_framework}--\Romannum{2}. We include the underlying mathematics and implementation details in Appendix\footnoteref{foot:Appendix}~Sec.~\ref{sec:shuffling_appendix}. 
    
    \smallskip
    \item[$\bullet$] {\bf Learning context via out-painting and in-painting.}
    To realize the self-supervised learning via out-painting, we generate an arbitrary number of windows of various sizes and aspect ratios, and superimpose them on top of each other, resulting in a single window of a complex shape. We then assign a random value to all pixels outside the window while retaining the original intensities for the pixels within. As for in-painting, we retain the original intensities outside the window and replace the intensity values of the inner pixels with a constant value.
    Unlike \cite{pathak2016context}, where in-painting is proposed as a proxy task by restoring only the patch central region, we restore the entire patch in the output. 
    Out-painting compels Models Genesis to learn global geometry and spatial layout of organs via extrapolating, while in-painting requires Models Genesis to appreciate local continuities of organs via interpolating.
    Examples of out-painting and in-painting are shown in \figurename~\ref{fig:unified_framework}--\Romannum{3} and \figurename~\ref{fig:unified_framework}--\Romannum{4}, respectively. More visualizations can be found in Appendix\footnote{\label{foot:Appendix}Appendix can be found in the full version at \href{http://www.cs.toronto.edu/~liang/Publications/ModelsGenesis/MICCAI_2019_Full.pdf}{tinyurl.com/ModelsGenesisFullVersion}}~Secs.~\ref{sec:outpainting_appendix}---\ref{sec:inpainting_appendix}.
\end{itemize}

\noindent Models Genesis have the following four unique properties:

\medskip
\noindent{\bf 1) Autodidactic---requiring no manual labeling.} Models Genesis are trained in a self-supervised manner with abundant unlabeled image datasets, demanding {\em zero} expert annotation effort. 
Consequently, Models Genesis are very different from traditional {\em supervised} transfer learning from ImageNet~\cite{shin2016deep,tajbakhsh2016convolutional}, which offers modest benefit to 3D medical imaging applications as well as that from the pre-trained models of NiftyNet\footnote{\label{foot:NiftyNet}NiftyNet Model Zoo: \href{https://github.com/NifTK/NiftyNetModelZoo}{https://github.com/NifTK/NiftyNetModelZoo}}, which is ineffective (see Sec.~\ref{sec:experiments} and Appendix\footnoteref{foot:Appendix} Sec.~\ref{sec:niftynet_appendix}) due to the small datasets and specific applications (\eg brain parcellation and organ segmentation) these models are trained for.

\medskip
\noindent{\bf 2) Eclectic---learning from multiple perspectives.} Our unified approach trains Models Genesis from multiple perspectives (appearance, texture, context, etc.), leading to more robust models across all target tasks, as evidenced in \tablename~\ref{tab:stand_alone}, where our unified approach is compared with our individual schemes. {This eclectic approach}, incorporating multiple tasks into a single image restoration task, empowers Models Genesis to learn more comprehensive representation.

\medskip
\noindent{\bf 3) Scalable---eliminating proxy-task-specific heads.}
Consolidated into a single image restoration task, our novel self-supervised schemes share the same encoder and decoder during training. Had each task required its own decoder, due to limited memory on GPUs, our framework would have failed to accommodate a large number of self-supervised tasks.
By unifying all tasks as a single image restoration task, any favorable transformation can be easily amended into our framework,
overcoming the scalability issue associated with multi-task learning~\cite{doersch2017multi}, where the network heads are subject to the specific proxy tasks.

\medskip
\noindent{\bf 4) Generic---yielding diverse applications.}
Models Genesis learn a general-purpose image representation that can be leveraged for a wide range of target tasks. Specifically, Models Genesis can be utilized to initialize the encoder for the target {\em classification} tasks and to initialize the encoder-decoder for the target {\em segmentation} tasks, while the existing self-supervised approaches are largely focused on providing encoder models only~\cite{jing2019self}. 
As shown in \tablename~\ref{tab:3D_target_tasks}, Models Genesis can be generalized across diseases (\eg nodule, embolism, tumor), organs (\eg lung, liver, brain), and modalities (\eg CT, X-ray, MRI), a generic behavior that sets us apart from all previous works in the literature where the  representation is learned via a specific self-supervised task; and thus lack generality. Such specific schemes include predicting the distance and 3D coordinates of two patches randomly sampled from a same brain~\cite{spitzer2018improving}, identifying whether two scans belong to the same person, predicting the level of vertebral bodies~\cite{jamaludin2017self}, and finally
the systematic study by Tajbakhsh~\etal\cite{tajbakhsh2019surrogate} where individualized self-supervised schemes are studied for a set of target tasks. 

\section{Experiments and Results}
\label{sec:experiments}

\begin{table}[t!b]
\scriptsize
\centering
\caption{
    Fine-tuning models from our Genesis Chest CT (3D) significantly outperforms learning from scratch in the five 3D target tasks ($p<0.05$).
    The cells checked by \xmark \ denote the properties that are different between the proxy  and target datasets. Our results show that our Genesis Chest CT generalizes across organs, diseases, datasets, and modalities. Footnotes show state-of-the-art performance for each target task.
}
\label{tab:3D_target_tasks}
    \begin{tabular}{p{0.06\linewidth}P{0.08\linewidth}P{0.08\linewidth}P{0.08\linewidth}P{0.09\linewidth}P{0.09\linewidth}P{0.17\linewidth}P{0.17\linewidth}P{0.1\linewidth}}
        \hline
        Task & Metric & Disease & Organ & Dataset & Modality & Scratch ($\%$) & Genesis ($\%$) & $p$-value \\
        \hline
        \texttt{NCC}$^{1}$ & AUC & & & & & 94.25$\pm$5.07 & \textbf{98.20$\pm$0.51} & 0.0180\\
        \texttt{NCS}$^{2}$ & IoU & & & & & 74.05$\pm$1.97 & \textbf{77.62$\pm$0.64} & 1.04$e$-4 \\
        \texttt{ECC}$^{3}$ & AUC & \xmark & & \xmark & & 79.99$\pm$8.06 & \textbf{88.04$\pm$1.40} & 0.0058 \\
        \texttt{LCS}$^{4}$ & IoU & \xmark & \xmark & \xmark & & 74.60$\pm$4.57 & \textbf{79.52$\pm$4.77} & 0.0361 \\
        \texttt{BMS}$^{5}$ & IoU & \xmark & \xmark & \xmark & \xmark & 90.16$\pm$0.41 & \textbf{90.60$\pm$0.20} & 0.0041 \\
        \hline
    \end{tabular}
    \begin{tablenotes}
        \scriptsize
        \item $^{1}$ \href{https://luna16.grand-challenge.org/results/}{LUNA winner} holds an official score of 0.968 vs. 0.971 (ours) 
        \item $^{2}$ {\href{https://ieeexplore.ieee.org/abstract/document/8363765}{Wu~\etal} holds a Dice of 74.05$\%$  vs. 75.86$\%\pm$0.90$\%$ (ours)}
        \item $^{3}$ \href{https://ieeexplore.ieee.org/document/8099989}{Zhou~\etal} holds an AUC of 87.06$\%$ vs. 88.04$\%\pm$1.40$\%$ (ours)
        \item $^{4}$ \href{https://competitions.codalab.org/competitions/17094#results}{LiTS winner} w/ postprocessing (PP) holds a Dice of 96.60$\%$ vs. 91.13$\%\pm$1.51$\%$ (ours  w/o  PP)
        \item $^{5}$ \href{https://www.smir.ch/BRATS/Start2013#leaderboardResults}{BraTS winner} w/ ensembling holds a Dice of 91.00$\%$ vs. 92.58$\%\pm$0.30$\%$ (ours w/o ensembling)
    \end{tablenotes}
\end{table}

\noindent{\bf Experiment protocol.} 
Our Genesis CT and Genesis X-ray are self-supervised pre-trained from 534 CT scans in LIDC-IDRI\footnoteref{foot:LIDC} and 77,074 X-rays in ChestX-ray8\footnoteref{foot:xray8}, respectively. 
The reason that we decided not to use all images in LIDC-IDRI and in ChestX-ray8 for training Models Genesis is to avoid test-image leaks between proxy and target tasks, so that we can confidently use the rest images solely for testing Models Genesis as well as the target models, although Models Genesis are trained from {\em only} unlabeled images, involving {\em no} annotation shipped with the datasets.
We evaluate Models Genesis in  seven medical imaging applications including 3D and 2D image classification and segmentation tasks  (codified as detailed in~\tablename~\ref{tab:terminology}).
For 3D applications in CT and MRI, we investigate the capability of both 2D slice-based solutions and 3D volume-based solutions; for 2D applications in X-ray and Ultrasound, we compare Models Genesis with random initialization and fine-tuning from ImageNet.  
3D U-Net architecture\footnote{\label{foot:VNet}3D U-Net Convolution Neural Network: \href{https://github.com/ellisdg/3DUnetCNN}{https://github.com/ellisdg/3DUnetCNN}} is used in five 3D applications; U-Net architecture with ResNet-18 encoder\footnote{\label{foot:SegmentationModels}Segmentation Models: \href{https://github.com/qubvel/segmentation_models}{https://github.com/qubvel/segmentation\_models}} is used in seven 2D applications. We utilize the L1-norm distance as the loss function in the image restoration tasks.
Performances of target image classification and segmentation tasks are measured by the AUC (Area Under the Curve) and IoU (Intersection over Union), respectively, through at least 10 trials.  
We report the performance metrics with mean and standard deviation and further present statistical analysis based on the independent two-sample $t$-test.

\medskip
\noindent{\bf Models Genesis outperform 3D models trained from scratch.} We evaluate the effectiveness of Genesis Chest CT in five distinct 3D medical target tasks. These target tasks are selected such that they show varying levels of semantic distance to the proxy task, as shown in~\tablename~\ref{tab:3D_target_tasks}, allowing us to investigate the transferability of Genesis Chest CT with respect to the domain distance.
\tablename~\ref{tab:3D_target_tasks} demonstrates that models fine-tuned from Genesis Chest CT consistently outperform their counterparts trained from scratch. Our statistical analysis show that the performance gain is significant for all the target tasks under study. Specifically, for \texttt{NCC} and \texttt{NCS} where the target and proxy tasks are in the same domain, initialization with Genesis Chest CT achieves 4 and 3 points increase in the AUC and IoU score, respectively, compared with training from scratch. 
For \texttt{ECC}, the target and proxy tasks are different in both the disease affecting the organ and the dataset itself; yet, Genesis Chest CT achieves a remarkable improvement over training from scratch, increasing the AUC by 8 points. Genesis Chest CT continues to yield significant IoU gain for \texttt{LCS} and \texttt{BMS} even though their domain distances with the proxy task are the widest. 
To our knowledge, we are the first to investigate cross-domain self-supervised learning in medical imaging. 
Given the fact that Genesis Chest CT is pre-trained on Check CT only, it is {\em remarkable} that our model can  generalize to different diseases, organs, datasets, and even modalities.

\begin{table}[t]
\scriptsize
\centering
\caption{
    Comparison between our unified framework and each of the suggested self-supervised schemes on five 3D target tasks. The statistical analyses is conducted between the top-2 models in each column {\zzred highlighted in red}. While there is no clear winner, our unified framework is more robust across all target tasks, yielding either the best result or comparable performance to the best model ($p>$ 0.05).
}
\label{tab:stand_alone}
\begin{tabular}{p{0.19\linewidth}P{0.15\linewidth}P{0.15\linewidth}P{0.15\linewidth}P{0.15\linewidth}P{0.15\linewidth}}
    \hline
    Approach & \texttt{NCC} ($\%$) & \texttt{NCS} ($\%$) & \texttt{ECC} ($\%$)& \texttt{LCS} ($\%$)& \texttt{BMS} ($\%$) \\
    \hline
    Scratch & 94.25$\pm$5.07 & 74.05$\pm$1.97 & 79.99$\pm$8.06 & 74.60$\pm$4.57 & 90.16$\pm$0.41 \\
    Distortion (ours) & 96.46$\pm$1.03 & \cellcolor{maroon!15}77.08$\pm$0.68 & \cellcolor{maroon!15}\textbf{88.04$\pm$1.40} & \cellcolor{maroon!15}79.08$\pm$4.26 & \cellcolor{maroon!15}\textbf{90.60$\pm$0.20} \\
    Painting (ours) & \cellcolor{maroon!15}\textbf{98.20$\pm$0.51} & 77.02$\pm$0.58 & 87.18$\pm$2.72 & 78.62$\pm$4.05 & 90.46$\pm$0.21 \\
    Unified (ours) & \cellcolor{maroon!15}97.90$\pm$0.57 & \cellcolor{maroon!15}\textbf{77.62$\pm$0.64} & \cellcolor{maroon!15}87.20$\pm$2.87 & \cellcolor{maroon!15}\textbf{79.52$\pm$4.77} & \cellcolor{maroon!15}90.59$\pm$0.21 \\
    \hline
    $p$-value & 0.0848 & 0.0520 & 0.2102 & 0.4249 & 0.4276 \\
    \hline
\end{tabular}
\end{table}

\medskip
\noindent{\bf Models Genesis consistently top any 2D approaches.} A common technique to handle limited data in medical imaging is to reformat 3D data into a 2D image representation followed by fine-tuning pre-trained ImageNet models~\cite{shin2016deep,tajbakhsh2016convolutional}. This approach increases the training examples by an order of magnitude, but it scarifies the 3D context. It is interesting to compare how Genesis Chest CT compares to this \textit{de facto} standard in 2D. For this purpose, we adopt the trained 2D models from an ImageNet pre-trained model\footnoteref{foot:SegmentationModels} for the tasks of \texttt{NCC}, \texttt{NCS}, and \texttt{ECC}. The 2D representation is obtained by extracting axial slices from volumetric datasets.
\tablename~\ref{tab:2D_3D_target_tasks} compares the results for 2D and 3D models. Note that the results for 3D models are identical to those reported in \tablename~\ref{tab:3D_target_tasks}. As evidenced by our statistical analyses, the 3D models trained from Genesis Chest CT significantly outperform the 2D models trained from ImageNet, achieving higher average performance and lower standard deviation (see~\tablename~\ref{tab:2D_3D_target_tasks} and  Appendix\footnoteref{foot:Appendix}~Sec.~\ref{sec:genesis_vs_imagenet_appendix}). However, the same conclusion does not apply to the  models trained from scratch{\zzred ---}3D scratch models outperform 2D scratch models in only two out of the three target tasks and also exhibit undesirably larger standard deviation. We attribute the mixed results of 3D scratch models to the larger number of model parameters and limited sample size in the target tasks, which together impede the full utilization of 3D context. In fact, the undesirable performance of the 3D scratch models highlights the effectiveness of Genesis Chest CT, which unlocks the power of 3D models for medical imaging.

\begin{table}[t]
\scriptsize
\centering
\caption{
    Comparison between 3D solutions and 2D slice-based solutions on three 3D target tasks. Training 3D models from scratch does not necessarily outperform the 2D counterparts (see \texttt{NCC}). However, training the same 3D models from Genesis Check CT outperforms ($p<0.05$) all 2D solutions, demonstrating the effectiveness of Genesis Chest CT in unlocking the power of 3D models.
}
    \begin{tabular}{P{0.06\linewidth}P{0.13\linewidth}P{0.13\linewidth}P{0.13\linewidth}P{0.000001\linewidth}P{0.13\linewidth}P{0.11\linewidth}P{0.13\linewidth}P{0.1\linewidth}}
    \hline
    \multirowcell{2}{Task} & \multicolumn{3}{c}{2D ($\%$)} & & \multicolumn{3}{c}{3D ($\%$)} & \multirowcell{2}{$p$-value$^{\dagger}$} \\
    \cline{2-4}\cline{6-8}
     & Scratch & ImageNet & Genesis & & Scratch & ImageNet & Genesis & \\
    \hline
    \texttt{NCC} & 96.03$\pm$0.86 & \cellcolor{maroon!15} 97.79$\pm$0.71 & 97.45$\pm$0.61 & & 94.25$\pm$5.07 & N/A & \cellcolor{maroon!15}\textbf{98.20$\pm$0.51} & 0.0213 \\
    \texttt{NCS} & 70.48$\pm$1.07 & \cellcolor{maroon!15}72.39$\pm$0.77 & 72.20$\pm$0.67 & & 74.05$\pm$1.97 & N/A & \cellcolor{maroon!15}\textbf{77.62$\pm$0.64} & $<$1$e$-8 \\
    \texttt{ECC} & 71.27$\pm$4.64 & \cellcolor{maroon!15}78.61$\pm$3.73 &  78.58$\pm$3.67 & & 79.99$\pm$8.06 & N/A & \cellcolor{maroon!15}\textbf{88.04$\pm$1.40} & 5.50$e$-4 \\
    \hline
    \end{tabular}
    \begin{tablenotes}
        \scriptsize
        \item $^{\dagger}$These $p$-values are calculated between our Models Genesis vs. the fine-tuning from ImageNet, which always offers the best performance (highlighted in red) for all three tasks in 2D.
    \end{tablenotes}
\label{tab:2D_3D_target_tasks}
\end{table}

\begin{figure}[t]
\footnotesize
\centering
    \includegraphics[width=0.496\linewidth]{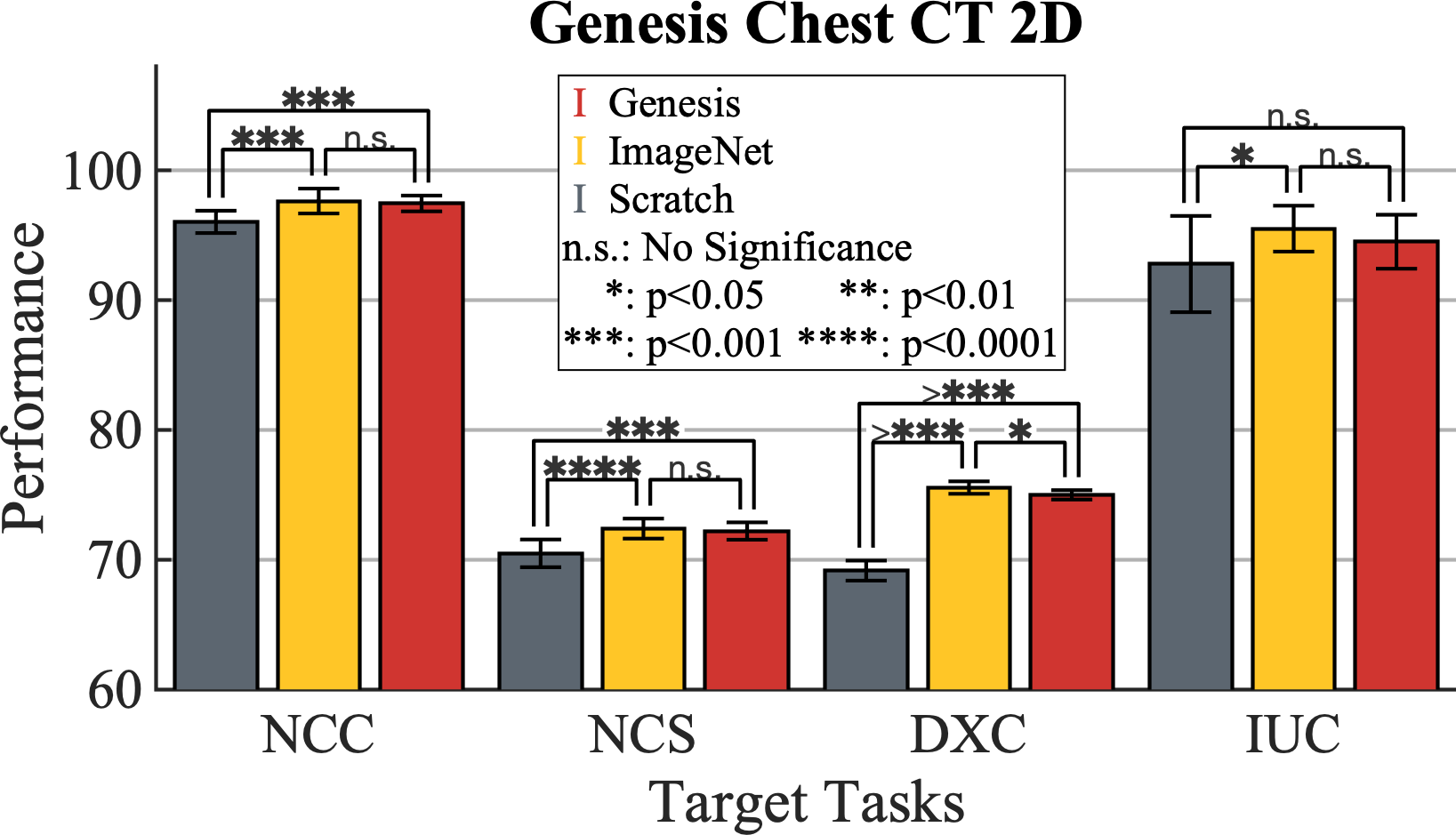}
    \includegraphics[width=0.496\linewidth]{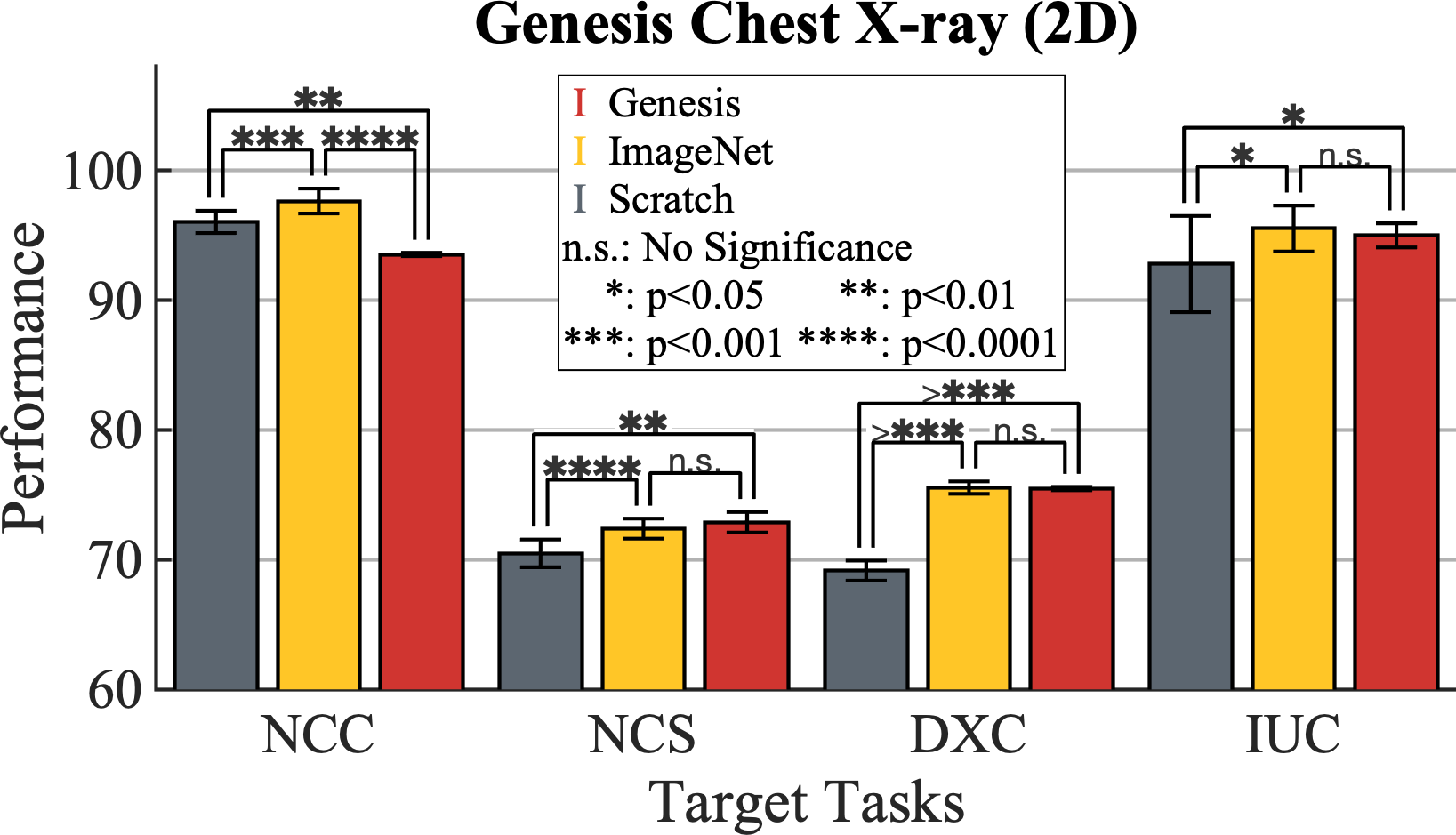}
\caption{
    Comparison of 2D solutions on four 2D target tasks. To investigate the same- and cross-domain transferability of Models Genesis, we have trained Genesis Chest CT 2D using 2D axial slices from LUNA dataset (left panel), and Genesis Chest X-ray (2D) trained using radiographs from ChestX-ray8 dataset (right panel). In same-domain target tasks (\texttt{NCC} and \texttt{NCS} in the left panel and \texttt{DXC} in the right panel), Models Genesis 2D outperform training from scratch and offer equivalent performance to fine-tuning from ImageNet. While in cross-domain target tasks (\texttt{DXC} and \texttt{IUC} in the left panel; \texttt{NCS} and \texttt{IUC} in the right panel), Models Genesis 2D also produce fairly robust performance.
}
\label{fig:2D_target_tasks}
\end{figure}

\medskip
\noindent{\bf Models Genesis (2D) offer equivalent performances to supervised pre-trained models.}
To compare our self-supervised approaches with those supervised pre-training from ImageNet~\cite{deng2009imagenet}, we deliberately downgrade our Models Genesis to 2D versions: Genesis Chest CT 2D and Genesis Chest X-ray (2D) (see visualization of Genesis 2D in Appendix\footnoteref{foot:Appendix}~Secs.~\ref{sec:genesis_chest_ct_appendix}---\ref{sec:genesis_chest_xray_appendix}).
The statistical analysis in \figurename~\ref{fig:2D_target_tasks} suggests that the downgraded Models Genesis 2D offer equivalent performance to state-of-the-art fine-tuning from ImageNet within modality, outperforming random initialization by a large margin, which is a significant achievement because ours comes at {\em zero} annotation cost.
Meanwhile, the downgraded Models Genesis 2D are fairly robust in cross-domain transfer learning, although they tend to underperform when domain distance is large, which suggests same-domain transfer learning should be preferred where possible in medical imaging.
For 3D applications, we also examine the effectiveness of fine-tuning from NiftyNet\footnoteref{foot:NiftyNet}, which is not designed for transfer learning but is the only available supervised pre-trained 3D model. Compared with training from scratch, fine-tuning NiftyNet suffers 3.37, 0.18, and 0.03 points decrease for \texttt{NCS}, \texttt{LCS}, and \texttt{BMS} tasks, respectively (detailed in Appendix\footnoteref{foot:Appendix} Sec.~\ref{sec:niftynet_appendix}), suggesting that strong supervision with limited annotated data cannot guarantee good transferability like ImageNet. Conversely, Models Genesis benefit from both large scale unlabeled datasets and dedicated proxy tasks which are essential for learning general-purpose visual representation.

\section{Conclusion and Future Work}

A key contribution of ours is a collection of \textit{generic source} models, nicknamed Models Genesis, built directly from {\em unlabeled} 3D image data with our novel unified self-supervised method, for generating powerful application-specific \textit{target} models through transfer learning.
While our empirical results are strong, surpassing state-of-the-art performances in most of the applications, an important future work is to extend our Models Genesis to modality-oriented models, such as Genesis MRI and Genesis Ultrasound, as well as organ-oriented models, such as Genesis Brain and Genesis Heart.
In fact, we envision that Models Genesis may serve as a primary source of transfer learning for 3D medical imaging applications, in particular, with limited annotated data. 
To benefit the research community, we make the development of Models Genesis open science, releasing our codes and models to the public, and inviting researchers around the world to contribute to this effort. We hope that our collective efforts will lead to the Holy Grail of Models Genesis, effective across diseases, organs, and modalities.

\medskip
\noindent {\bf Acknowledgments:} This research has been supported partially by ASU and Mayo Clinic through a Seed Grant and an Innovation Grant, and partially by NIH under Award Number R01HL128785. The content is solely the responsibility of the authors and does not necessarily represent the official views of NIH.

\title{NablaNet: A Nested Neural Network Architecture\\for Medical Image Segmentation}
\title{$\nabla$-Net: A Nested Ensemble of Neural Networks for Medical Image Segmentation}
\title{$\nabla$-Net: A Upside-down Triangulated Neural Network Architecture for Medical Image Segmentation}
\title{$\nabla$-Net: A Nested Neural Network Architecture for Medical Image Segmentation}

\title{Supplementary Material\\Models Genesis: Generic Autodidactic Models for 3D Medical Image Analysis}

%
\titlerunning{Supplementary Material for Models Genesis}
%
\author{
Zongwei Zhou\inst{1} \and
Vatsal Sodha\inst{1} \and
Md Mahfuzur Rahman Siddiquee\inst{1} \and \\
Ruibin Feng\inst{1} \and
Nima Tajbakhsh\inst{1} \and
Michael B. Gotway\inst{2} \and
Jianming Liang\inst{1}
}
%
\authorrunning{Z. Zhou et al.}
%
\institute{
Arizona State University, Scottsdale, AZ 85259 USA \\
\email{\{zongweiz,vasodha,mrahmans,rfeng12,ntajbakh,jianming.liang\}@asu.edu} \and 
Mayo Clinic, Scottsdale, AZ 85259 USA \\
\email{Gotway.Michael@mayo.edu}
}
\maketitle              
\setcounter{footnote}{0}

\setcounter{figure}{2}

\begin{abstract}
{\zzred
    This document provides supplementary material for the paper entitled ``Models Genesis: Generic Autodidactic Models for 3D Medical Image Analysis''.
    The supplementary material is organized as follows.
    In Sec.~\ref{sec:method_appendix}, we begin with a brief overview of Models Genesis.
    Secs.~\ref{sec:nonlinear_appendix}---\ref{sec:inpainting_appendix} describe at length the detailed implementation and illustration of four individual transformations.
    Secs.~\ref{sec:genesis_chest_ct_appendix}---\ref{sec:genesis_chest_xray_appendix} contain a qualitative visualization on the pre-trained Genesis CT and Genesis X-ray for both same- and cross-domain image restoration.
    Secs.~\ref{sec:genesis_vs_imagenet_appendix}---\ref{sec:niftynet_appendix} present the transfer learning results of Models ImageNet, NiftyNet, and our Models Genesis in various target tasks.
}
\end{abstract}

\begin{figure}[!h]
\centering
\includegraphics[width=1.0\linewidth]{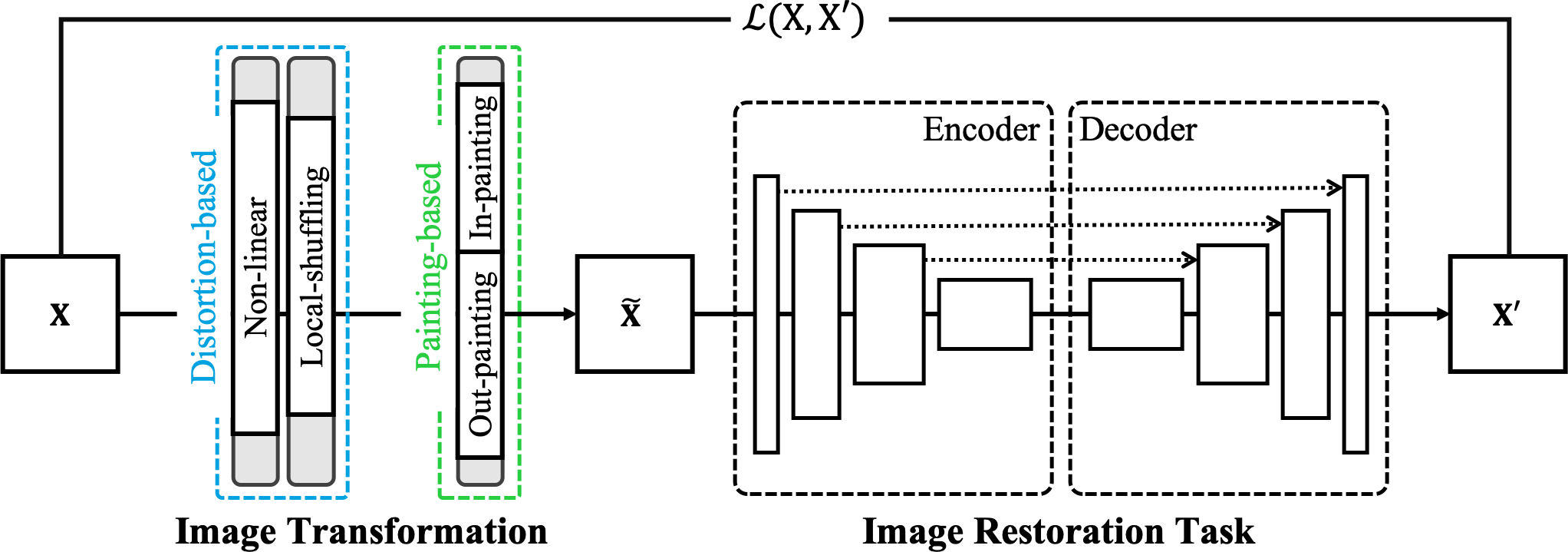}
\caption{
    Overview of our unified self-supervised learning framework. Given an image, we first extract patches $\text{X}$ of arbitrary sizes from random locations and then apply the transformations on them as mentioned in \figurename~\ref{fig:unified_framework_appendix}. Models Genesis learns the visual representation by restoring the original patches $\text{X}$ from the transformed ones $\tilde{\text{X}}$.
} 
\label{fig:architecture_appendix}
\end{figure}

\appendix

\newpage
\section{Models Genesis}
\label{sec:method_appendix}

\begin{figure}[!h]
\centering
\includegraphics[width=1\linewidth]{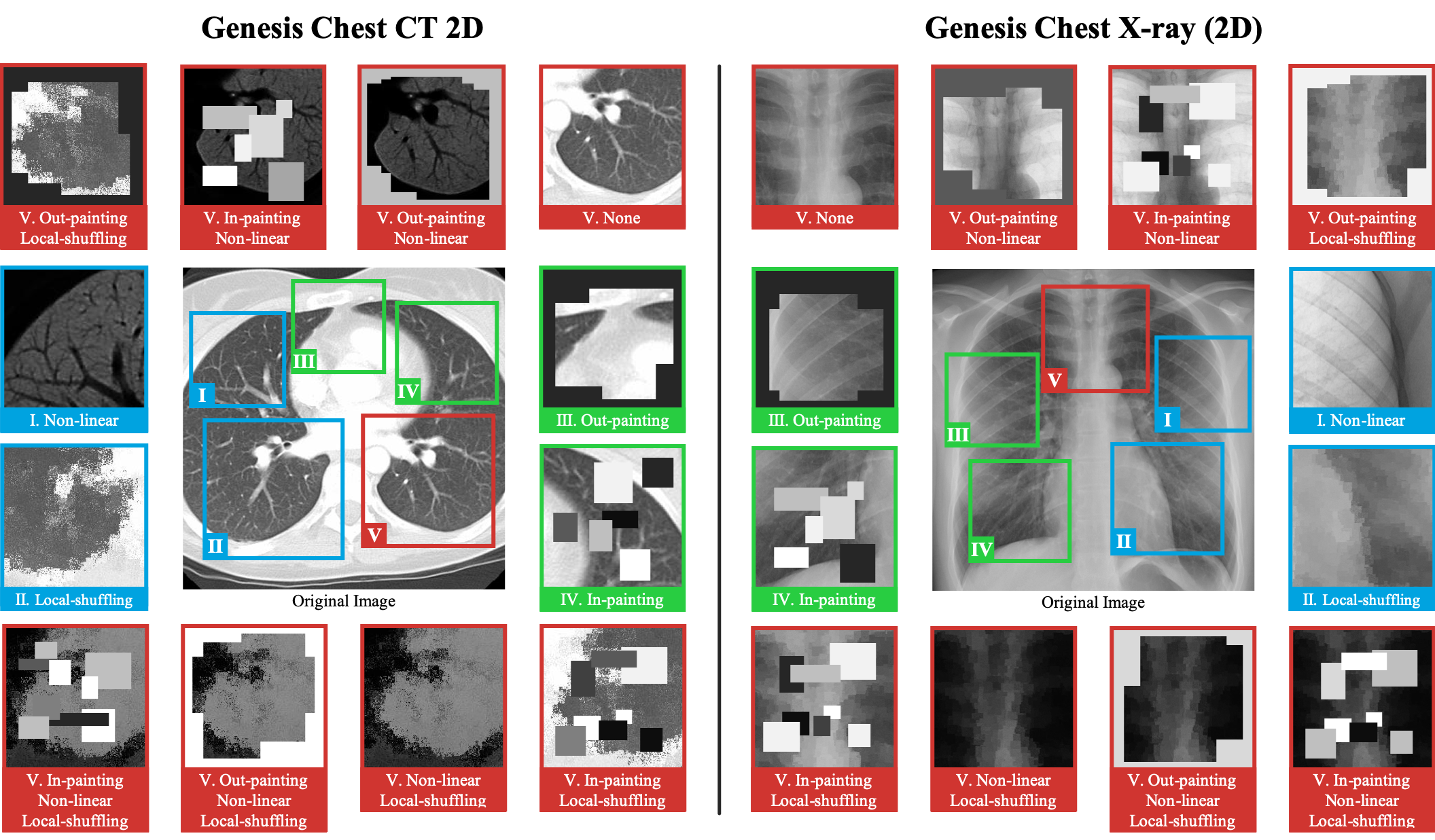}
\caption{
    [Better viewed on-line in color and zoomed in for details] 
    Our novel unified self-supervised learning framework aims to learn general-purpose visual representation by recovering original image patches from their transformed ones.
    We have designed four individual transformations:
    I) non-linear (see Sec.~\ref{sec:nonlinear_appendix}), 
    II) local-shuffling (see Sec.~\ref{sec:shuffling_appendix}), 
    III) out-painting (see Sec.~\ref{sec:outpainting_appendix}), and 
    IV) in-painting (see Sec.~\ref{sec:inpainting_appendix}). 
    We have provided examples of the transformed images for Genesis Chest CT (left) and Genesis Chest X-ray (right). 
    For simplicity and clarity, we illustrate our idea on a 2D CT slice and a 2D X-ray image, but our Genesis Chest CT is trained using 3D Check CT images directly. 
    Each transformation is independently applied to a patch with a predefined probability, while out-painting and in-painting are considered mutually exclusive.
    Therefore, in addition to the four original individual transformations, this process yields eight more transformations framed in red, including one identity mapping (\ie V: none, meaning none of the four individual transformations is selected) and seven combined transformations as indicated under each patch framed in red. For clarity, we further define a {\em training scheme} as the process that transforms patches with any of the twelve aforementioned transformations and trains a model to restore the original patches from the transformed ones. For convenience, we refer to an {\em individual training scheme} as the scheme using one particular individual transformation. Finally, our unified learning framework utilizes all possible transformations randomly with pre-defined probabilities and trains a model to restore the original patches from the ones undergone any possible transformations.
}
\label{fig:unified_framework_appendix}
\end{figure}

As shown in~\figurename~\ref{fig:architecture_appendix}, our proposed self-supervised learning framework consists of two components: image transformation (illustrated in \figurename~\ref{fig:unified_framework_appendix}) and image restoration, where Models Genesis, taking an encoder-decoder architecture, are trained by restoring original patch $\text{X}$ from transformed patch $\tilde{\text{X}}$, aiming to learn common visual representation that is transferable and generalizable across diseases, organs and, modalities and thus yield high-performance target models. From this work, we have concluded: 
\begin{enumerate}
      \item Models Genesis significantly outperform learning from scratch in all five target 3D applications covering both segmentation and classification. More importantly, learning a model from scratch {\em simply in 3D} may not necessarily yield performance better than transfer learning from ImageNet in 2D, but Models Genesis consistently top any 2D approaches including fine-tuning from ImageNet~\cite{krizhevsky2012imagenet} as well as fine-tuning our 2D Models Genesis, confirming the importance of 3D anatomical information and significance of our Models Genesis for 3D medical imaging. 
      \item Despite the outstanding performance of Models Genesis, a large, strongly annotated dataset for medical image analysis like ImageNet~\cite{deng2009imagenet_appendix} for computer vision is still highly demanded. In computer vision, {\zzred at the time this paper is written,} no self-supervised learning method outperforms fine-tuning models pre-trained from ImageNet~\cite{jing2019self_appendix,chen2019self,kolesnikov2019revisiting}. One of our goals of developing Models Genesis is to help create such a large, strongly annotated dataset for medical image analysis, because based on a small set of expert annotations, models fine-tuned from Models Genesis will be able to help quickly generate initial rough annotations of unlabeled images for expert review, thus reducing the annotation efforts and accelerating the creation of a large, strongly annotated, medical ImageNet. In summary, Models Genesis are not designed to replace such a large, strongly annotated dataset for medical image analysis like ImageNet for computer vision, but rather helping create one.
      \item Same-domain transfer learning is always preferred whenever possible. Same-domain transfer learning strikes as a preferred choice in terms of performance; therefore, as our future work, we continue training modality-oriented models, including Genesis CT, Genesis MRI, Genesis X-ray, and Genesis Ultrasound, as well as organ-oriented models, including Genesis Brain, Genesis Lung, Genesis Heart, and Genesis Liver. 
      \item Cross-domain transfer learning is the Holy Grail. Retrieving a large number of unlabeled images from a PACS system requires an IRB approval, often a long process; the retrieved images must be de-identified; organizing the de-identified images in a way suitable for deep learning is tedious and laborious. Therefore, large quantities of unlabeled datasets may not be readily available to many target domains. Evidenced by our results in \tablename~\ref{tab:3D_target_tasks} and \figurename~\ref{fig:2D_target_tasks}, Models Genesis have a great potential for cross-domain transfer learning; particularly, distortion-based approaches take advantage of relative intensity values (in all modalities) to learn shapes and appearances of various organs. Therefore, as our future work, we will be focusing on methods that generalize well in cross-domain transfer learning. Building the Holy Grail of Models Genesis, effective across diseases, organs, and modalities, takes a village. {\zzred As a result, we make the development of Models Genesis open science, inviting researchers around the world to join this effort. All pre-trained Models Genesis will be made public at \href{https://github.com/MrGiovanni/ModelsGenesis}{https://github.com/MrGiovanni/ModelsGenesis}.}
\end{enumerate}

\newpage
\section{Non-linear Intensity Transformation Visualization}
\label{sec:nonlinear_appendix}

\begin{figure}[h]
\centering
\includegraphics[width=1.0\linewidth]{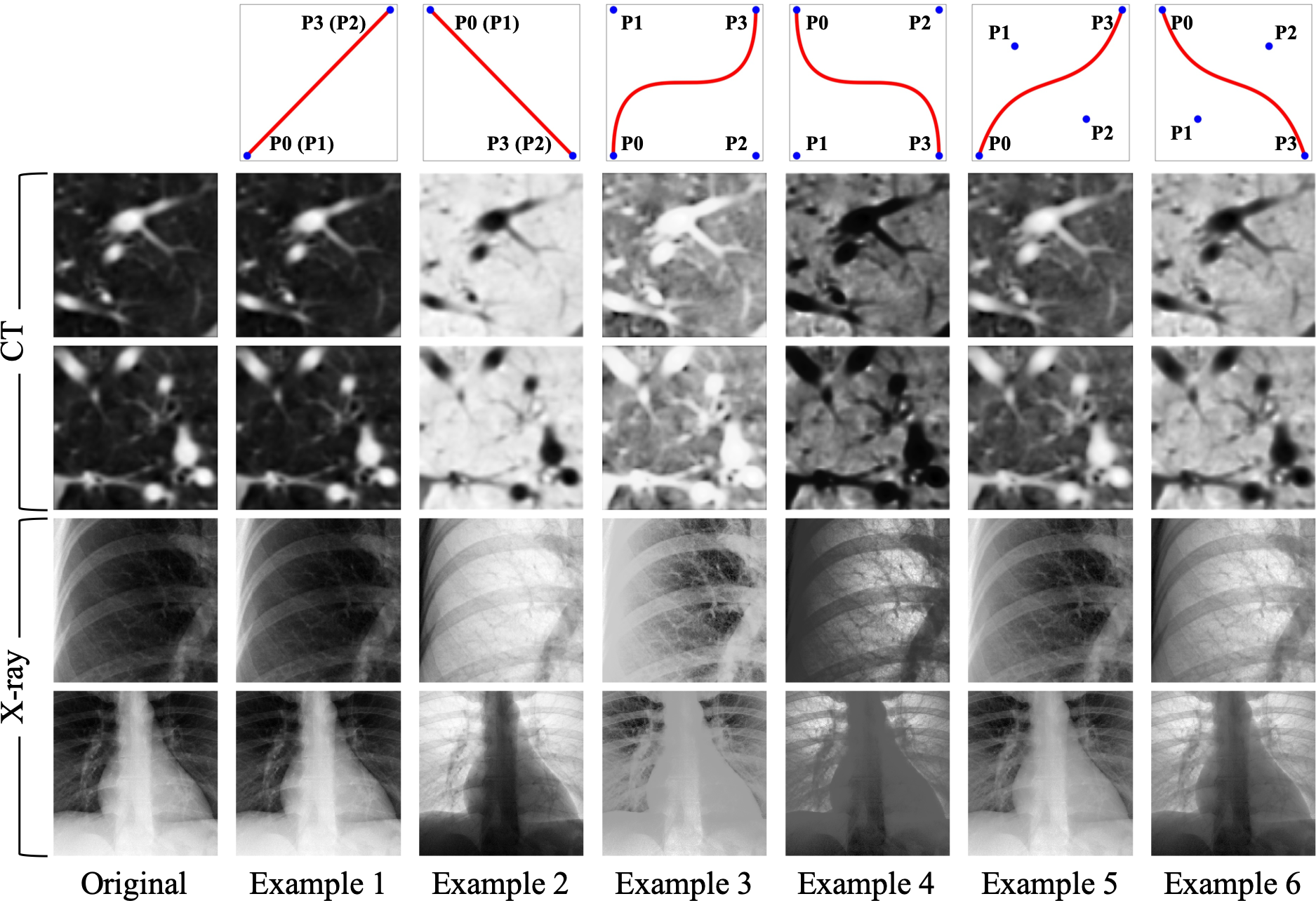}
\caption{
    We adopt non-linear intensity transformation as a new training scheme for self-supervised learning, which allows the model to learn the absolute or relative appearance of structures.
    Illustration of various non-linear intensity transformations (Examples~1---6) for CT (Rows~2---3) and X-ray (Rows~4---5) images is provided. We utilize four control points ($P_0$---$P_3$) in Eq.~\ref{eq:nonlinear_appendix} to modify the shape of the transformation function (Row 1). Notice that, when $P_0 = P_1$ and $P_2 = P_3$ the transformation function is a linear function (shown in Examples 1---2). Besides, we set $P_0 = (0,0)$ and $P_3 = (1,1)$ to get an increasing function (shown in Examples 1, 3, and 5) and the opposite to get a decreasing function (shown in Examples 2, 4, and 6). The control points are randomly generated for more variances.
}
\label{fig:non-linear_appendix}
\end{figure}

We propose a novel self-supervised training scheme based on non-linear translation, with which the model learns to restore the intensity values of the input image transformed with a set of non-linear functions. The rationale is that the absolute intensity values (\ie Hounsfield Units) in CT scans or relative intensity values in other imaging modalities convey important information about the underlying structures and organs~\cite{buzug2011computed,forbes2012human}. Hence, this training scheme enables the model to learn the appearance of the anatomic structures present in the images. In order to keep the appearance of the anatomic structures perceivable, we keep the non-linear intensity transformation function \textit{monotonic}, allowing pixels of different values to be assigned with new distinct values. To realize this idea, we use B{\'e}zier Curve~\cite{mortenson1999mathematics}, a smooth and monotonous transformation function, which is generated from two end points ($P_0$ and $P_3$) and two control points ($P_1$ and $P_2$), defined as: 
\begin{equation}
B(t)=(1-t)^3P_0+3(1-t)^2tP_1+3(1-t)t^2P_2+t^3P_3,\ t\in [0,1], 
\label{eq:nonlinear_appendix}
\end{equation}
where $t$ is a fractional value along the length of the line. In~\figurename~\ref{fig:non-linear_appendix}, we illustrate the original patches (the left-most column) and the transformed patches of 2D CT and X-rays based on different transformation functions. The corresponding transformation functions are shown in the top row. 
In order to apply the transformation functions on CT images, we first clip the HU values to a range of $[-1000, 1000]$ and then normalize to $[0, 1]$ for each of the CT image slices. In contrast, the X-ray images are directly normalized to $[0, 1]$ without intensity clipping.

\newpage
\section{Local Pixel Shuffling Visualization}
\label{sec:shuffling_appendix}

\begin{figure}[h]
\centering
\includegraphics[width=1.0\linewidth]{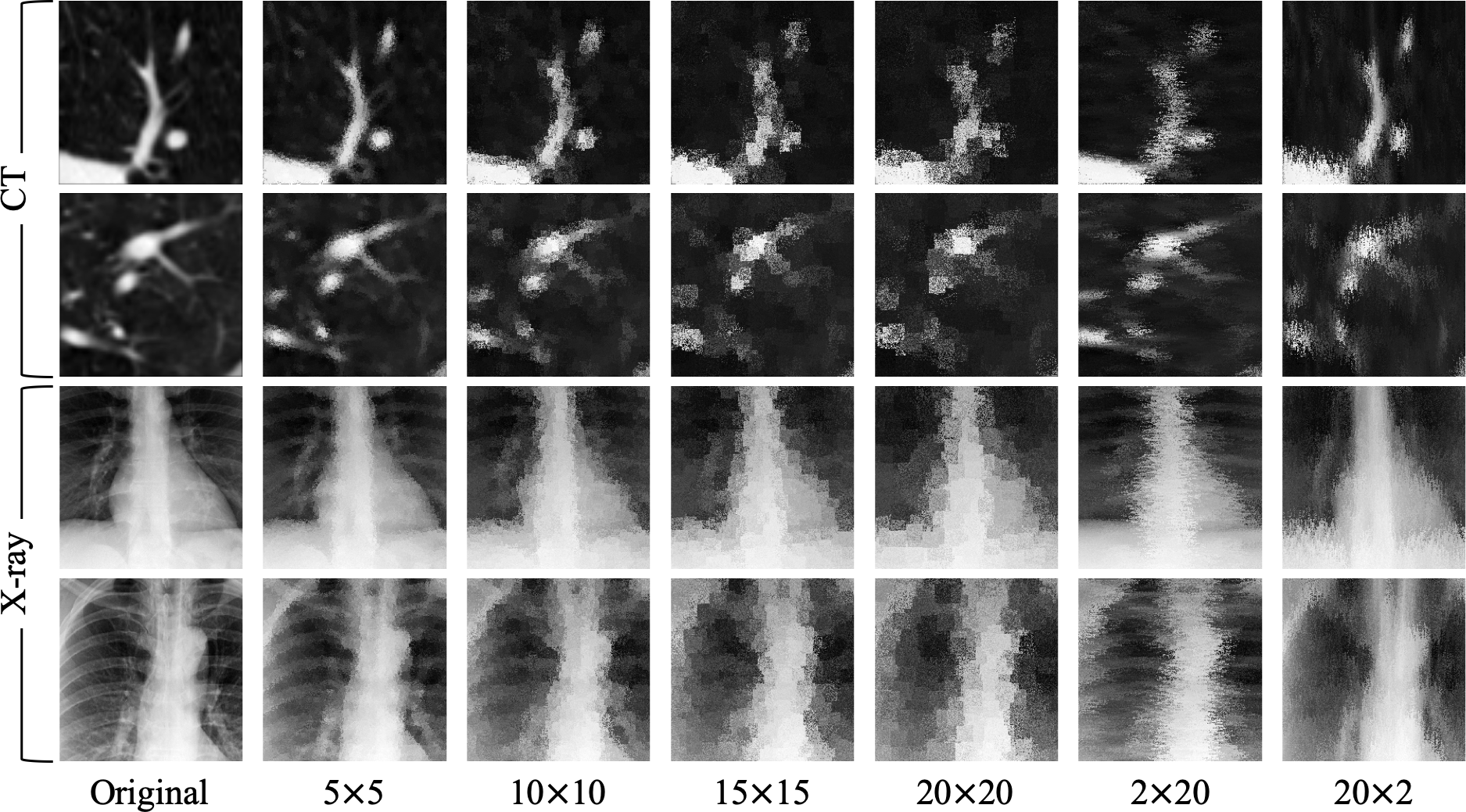}
\caption{
    We adopt local pixel shuffling as a new training scheme for self-supervised learning, which allows the model to learn the global geometry and spatial layout of organs as well as the local shape and texture of organs.
    Illustration of local pixel shuffling using multiple window sizes (Columns~2---7) applied on CT (Rows~1---2) and X-ray (Rows~3---4) images is provided. When $5 \times 5$ window is applied, the shapes are largely maintained; while the ribs are mostly invisible for window size equal to $20 \times 20$. Besides, various aspect ratios of windows also impose more local variances in different directions. Taking the restored X-ray patches in the last two columns as examples, a window size with $h\ll w$ (Column 6) distorts the boundary of the spine while preserving the overall presence of the ribs. On the other hand, when $h\gg w$ (Column 7), the ribs are hardly visible but the width of spine and heart is barely changed.
}
\label{fig:shuffling_appendix}
\end{figure}

We propose local pixel shuffling to enrich local variations of a patch without dramatically compromising its global structures, which encourages the model to learn the shapes and boundaries of the objects as well as the relative layout of different parts of the objects. To be specific, for each input patch, we randomly select 1,000 windows from the patch and then shuffle the pixels inside each window sequentially. 
Mathematically, let us consider a small window $\mathbf{W}$ with the size of $m\times n$.
The local-shuffling acts on each window and can be formulated as
\begin{equation}
\tilde{\mathbf{W}}=\mathbf{P}\times\mathbf{W}\times\mathbf{P}',
\end{equation}
where $\tilde{\mathbf{W}}$ is the transformed window, $\mathbf{P}$ and $\mathbf{P}'$ denote permutation metrics with the size of $m\times m$ and $n\times n$, respectively. 
Pre-multiplying $\mathbf{W}$ with $\mathbf{P}$ permutes the rows of the window $\mathbf{W}$, whereas post-multiplying $\mathbf{W}$ with $\mathbf{P}'$ results in the permutation of the columns of the window $\mathbf{W}$.
In practice, we keep the window sizes smaller than the receptive field of the network, so that the network can learn a more robust visual representation by ``resetting'' the original pixel positions.
To facilitate the understanding, we have explored the local-shuffling transformation of varying window sizes and illustrated them along with the original patches.
The window sizes can control the degree of distortion. 
As shown in~\figurename~\ref{fig:shuffling_appendix}, local-shuffling within an extent keeps the objects perceivable, it will benefit the deep neural network in learning invariant visual representations by restoring the original patches.
Unlike de-noising~\cite{vincent2010stacked} and in-painting~\cite{pathak2016context_appendix,iizuka2017globally}, our local-shuffling transformation does not intend to replace the pixel values with noise, which therefore preserves the identical global distributions to the original patch. 

\newpage
\section{Out-painting Visualization}
\label{sec:outpainting_appendix}
\begin{figure}[!h]
\centering
\includegraphics[width=1.0\linewidth]{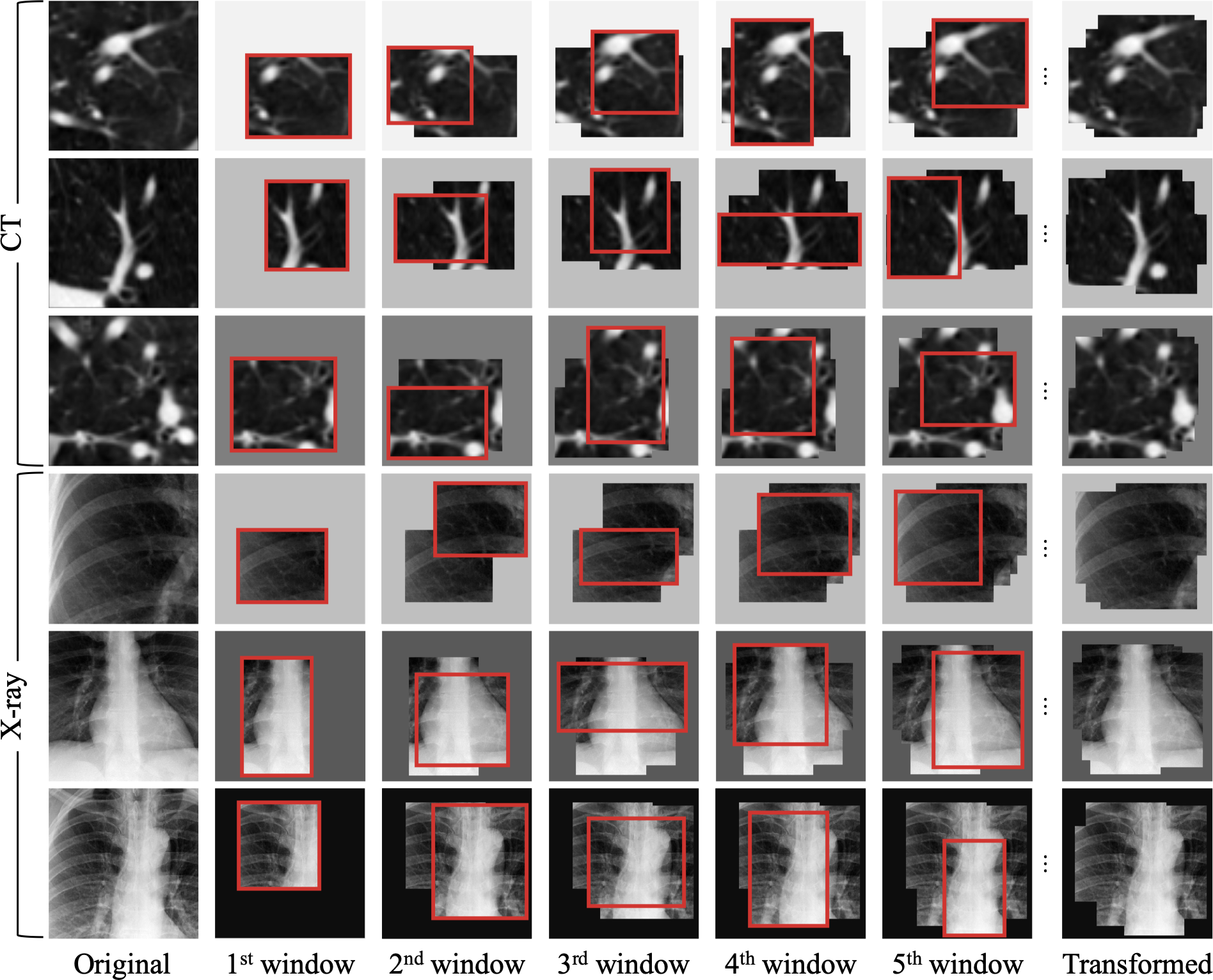}
\caption{
    We adopt out-painting as a new training scheme for self-supervised learning, which allows the model to learn the {\em global} geometry and spatial layout of organs.
    Illustration of the transformation for out-painting using various window sizes in CT (Rows~1---3) and X-ray (Rows~4---6) images is provided.
    The first and last columns denote the original patches and the final transformed patches, respectively. 
    From Column 2 to Column 6, we generate a new window (red framed) and merge it with the existing ones. 
    Moreover, to prevent the task to be too difficult or even unsolvable, we limit the masked surrounding region less than $1/4$ of the whole patch.
}
\label{fig:outpainting_appendix}
\end{figure}

We devise out-painting as a new training scheme for self-supervised learning, which allows the network to learn  {\em global} geometry and spatial layout of organs in medical images by extrapolation.
To realize it, we generate an arbitrary number ($\leq 10$) of windows with various sizes and aspect ratios, and superimpose them on top of each other, resulting in a single window of a complex shape.
When applying this merged window to a patch, we leave the patch region inside the window exposed and mask its surrounding with a random number.
We have illustrated this process step by step in~\figurename~\ref{fig:outpainting_appendix}. 

\newpage
\section{In-painting Visualization}
\label{sec:inpainting_appendix}

\begin{figure}[!h]
\centering
\includegraphics[width=1.0\linewidth]{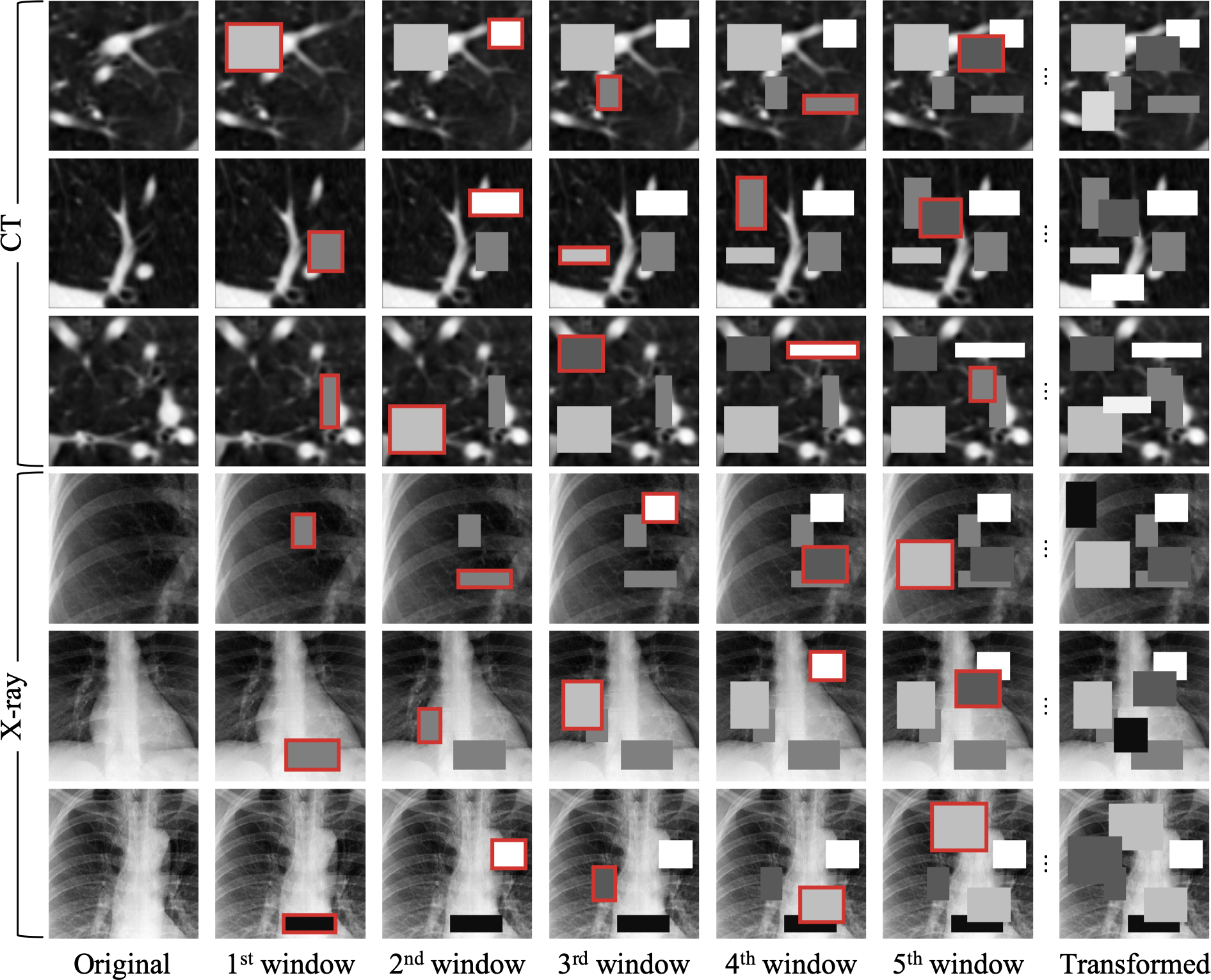}
\caption{
    We adopt in-painting as a new training scheme for self-supervised learning, which allows the model to learn {\em local} shape and texture of organs in medical images via interpolation. 
    The final transformed patches (Column 7) are obtained by iteratively superimposing a window of random size and aspect ratio, filled with a random number, to the original patches (Column 1). Columns 2---6 illustrate this process step by step. Similar to out-painting, the masked areas are also limited to be less than $1/4$ of the whole patch, in order to keep the task reasonably difficult.
}
\label{fig:inpainting_appendix}
\end{figure}


\newpage
\section{Genesis Chest CT}
\label{sec:genesis_chest_ct_appendix}

\begin{figure}[!h]
\centering
\includegraphics[width=1.0\linewidth]{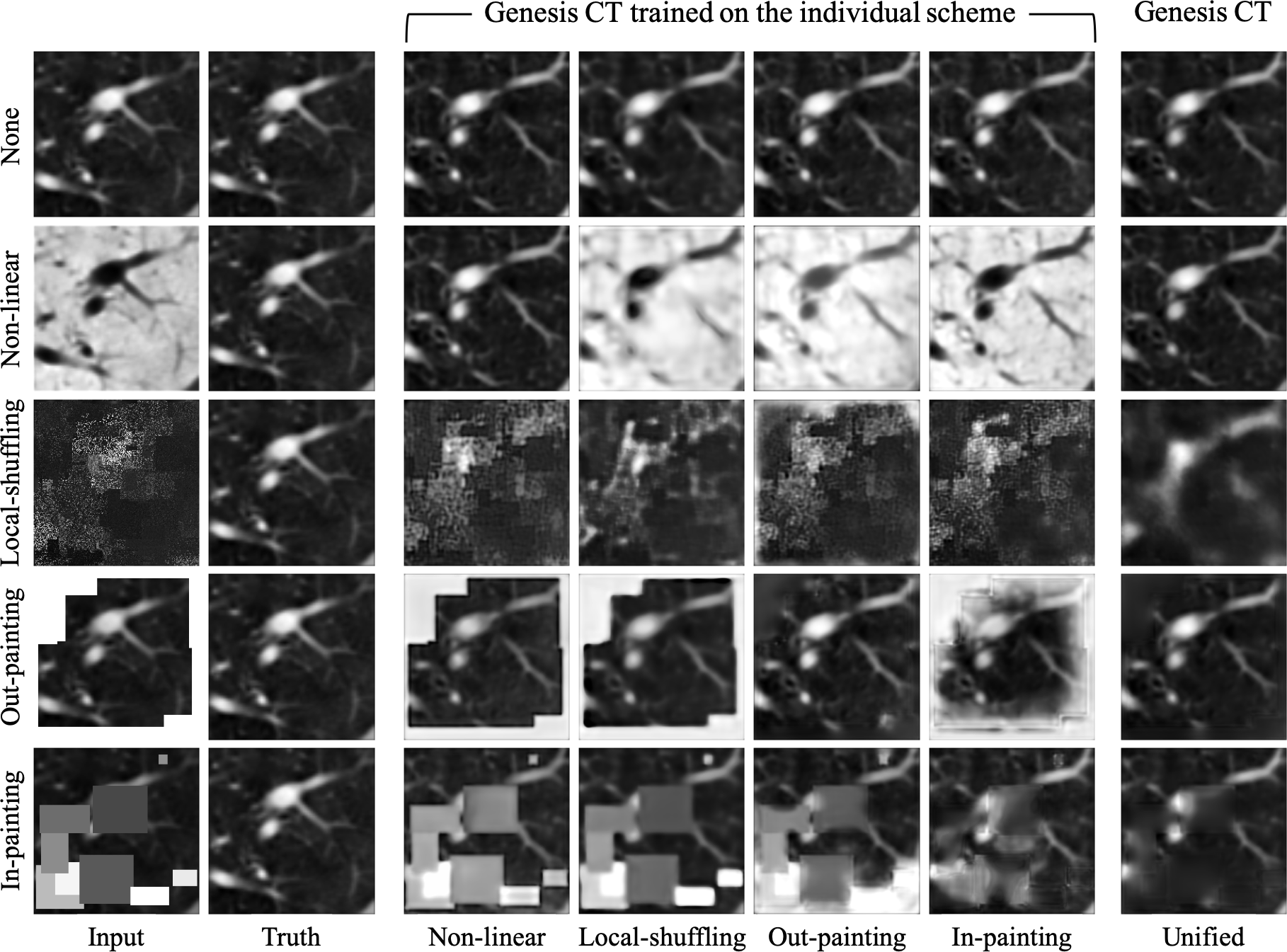}
\caption{
    Qualitative assessment of CT image restoration quality using Models Genesis trained with different training schemes, including the unified framework and four individual training schemes. LIDC-IDRI~\cite{armato2011lung} is used for both training and testing. We test these models with transformed patches that have undergone four individual transformations as well as the identity mapping (\ie no transformation). First of all, it can be seen the models trained with single-transformation-based schemes fail to handle other transformations. Taking non-linear transformation (Row 2) as an example, any individual training scheme besides non-linear transformation itself cannot invert the pixel intensity from transformed whitish to the original blackish. As expected, the model trained with the unified framework successfully restores original images from various transformations. Second, the model trained with the unified framework shows its superior to other models even if they are trained with and tested on the same transformation. For example, in local-shuffling case (Row 3), the patch recovered from the local-shuffling pre-trained model (Column 4) is noisy and lacks texture. However, the model trained with the unified framework (Column 7) generates a patch with more underlying structures, which demonstrates that learning with augmented tasks can even improve the performance on each individual tasks. These observations suggest that the model trained with the proposed unified self-supervised learning framework can successfully learn general anatomical structures and yield promising transferability on different target tasks.
} 
\label{fig:genesis_chest_ct_within_modality_appendix_1}
\end{figure}

\newpage
\begin{figure}[!h]
\centering
\includegraphics[width=1.0\linewidth]{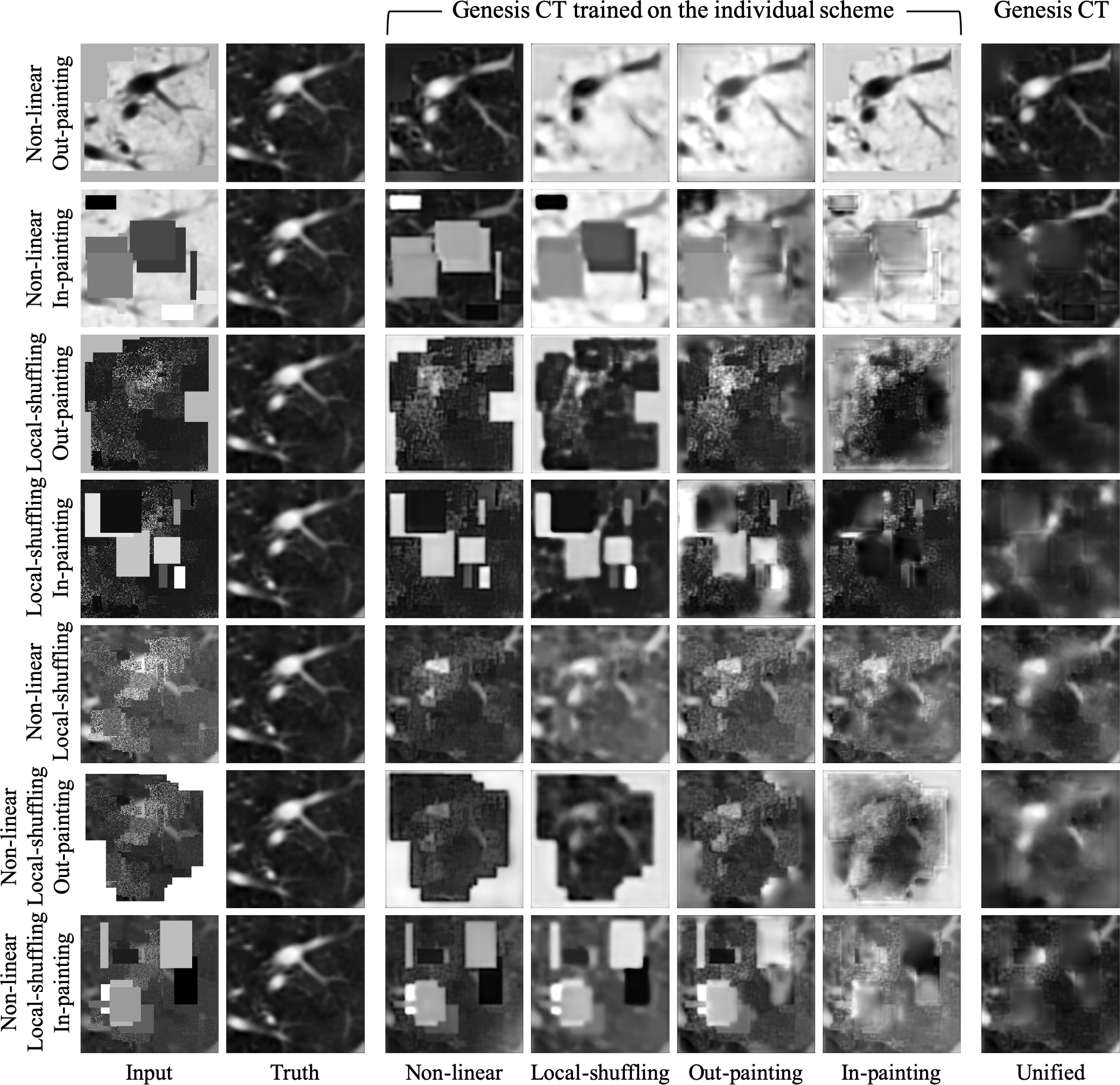}
\caption{
    Continued from~\figurename~\ref{fig:genesis_chest_ct_within_modality_appendix_1}. To further test our models, we show the restoration results on CT slices undergone seven different combined transformations. As expected, the model trained with our unified self-supervised framework significantly outperforms models trained with individual training schemes, further demonstrating the effectiveness of the proposed unified training framework as well as the pre-trained Models Genesis.
} 
\label{fig:genesis_chest_ct_within_modality_appendix_2}
\end{figure}

\newpage
\begin{figure}[!h]
\centering
\includegraphics[width=1.0\linewidth]{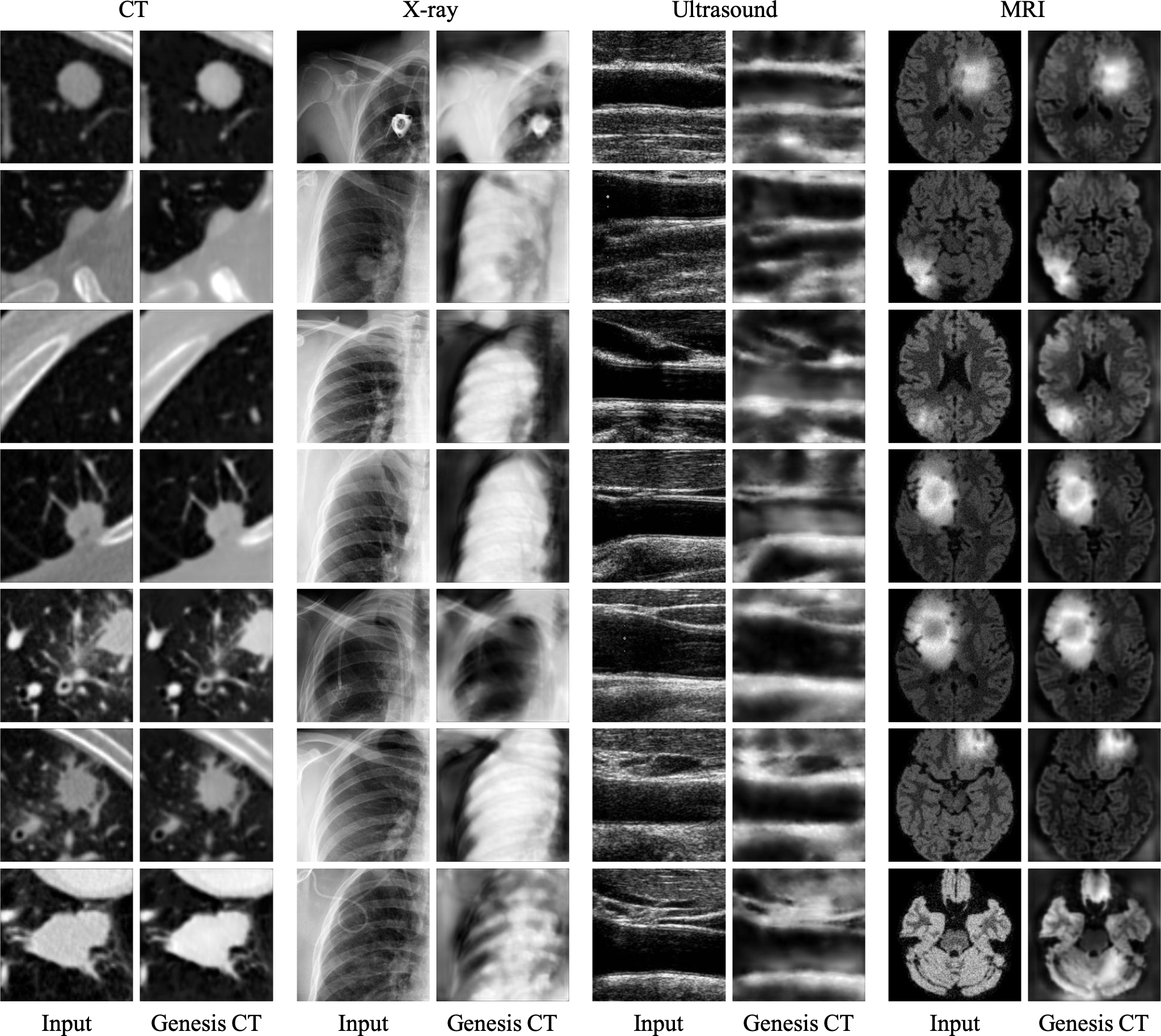}
\caption{
    Qualitative assessment of image restoration quality by Genesis Chest CT across dataset, organ, and modality is visualized. Genesis Chest CT is trained on LIDC-IDRI (CT)~\cite{armato2011lung} via our unified self-supervised training framework. For testing, we use the pre-trained model to directly restore images from LIDC-IDRI (CT), ChestX-ray8 (X-ray)~\cite{wang2017chestx}, CIMT (Ultrasound)~\cite{hurst10,zhou2019integrating}, and BraTS (MRI)~\cite{menze2015multimodal}. Though the model is only trained on CT data, it can largely maintain the texture and structures during restoration not only in the same modality (CT), but also in different modalities including X-ray, Ultrasound, and MRI, suggesting that Genesis Chest CT is transferable across datasets, organs, and modalities. Besides, we notice that the restoration quality is also consistent with the results of Genesis Chest CT on target tasks (see Fig.~\ref{fig:2D_target_tasks}). For example, compared to cross-modality performance, Genesis Chest CT yields better performance in CT for both restoration and target tasks (\ie \texttt{NCC} and \texttt{NCS}). Moreover, the relative lower restoration quality of ultrasound images may explain the relative lower target performance of Genesis Chest CT on \texttt{IUC} (see Fig.~\ref{fig:2D_target_tasks}). Finally, by comparing the performance of Genesis Chest CT in various modalities, we find out that a model pre-trained in the same domain is still preferred whenever possible. Thereby, we will continue developing modality-oriented models including Genesis MRI and Genesis Ultrasound.
}
\label{fig:genesis_chest_ct_cross_modality_appendix}
\end{figure}

\newpage
\section{Genesis Chest X-ray}
\label{sec:genesis_chest_xray_appendix}
\begin{figure}[!h]
\centering
\includegraphics[width=1.0\linewidth]{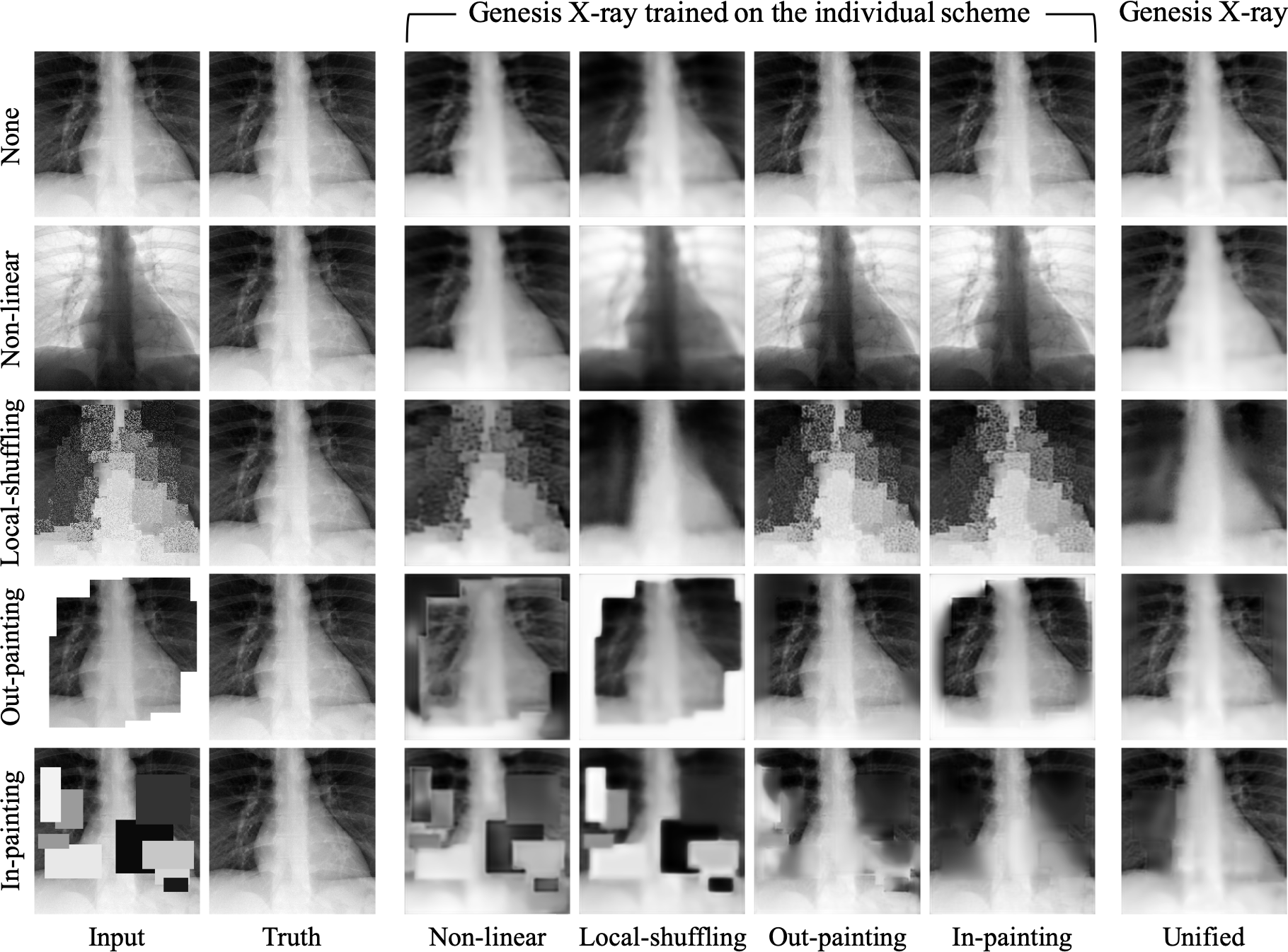}
\caption{
    Qualitative comparisons of Genesis Chest X-ray trained with unified self-supervised framework and four individual training schemes. We train and test all five models on ChestX-ray8~\cite{wang2017chestx} where transformed patches (Column 1) undergo one of the four transformations (Rows 2---5) as well as an identity mapping (Row 1). It is clear from the figure that the models trained with a single transformation fail to handle other transformations. For example, considering the training scheme based on in-painting (Row 5), models trained on individual training schemes fail to in-paint the masked region except for the in-painting-trained model (Column 6). However, the model trained with the unified framework (Column 7) handles all of the transformations and generates patches fairly close to the ground truths. These observations suggest that Models Genesis trained with proposed unified self-supervised learning framework learns general anatomical structures better, yielding high-performance target models through transfer learning.
}
\label{fig:genesis_chest_xray_within_modality_appendix_1}
\end{figure}

\newpage
\begin{figure}[!h]
\centering
\includegraphics[width=1.0\linewidth]{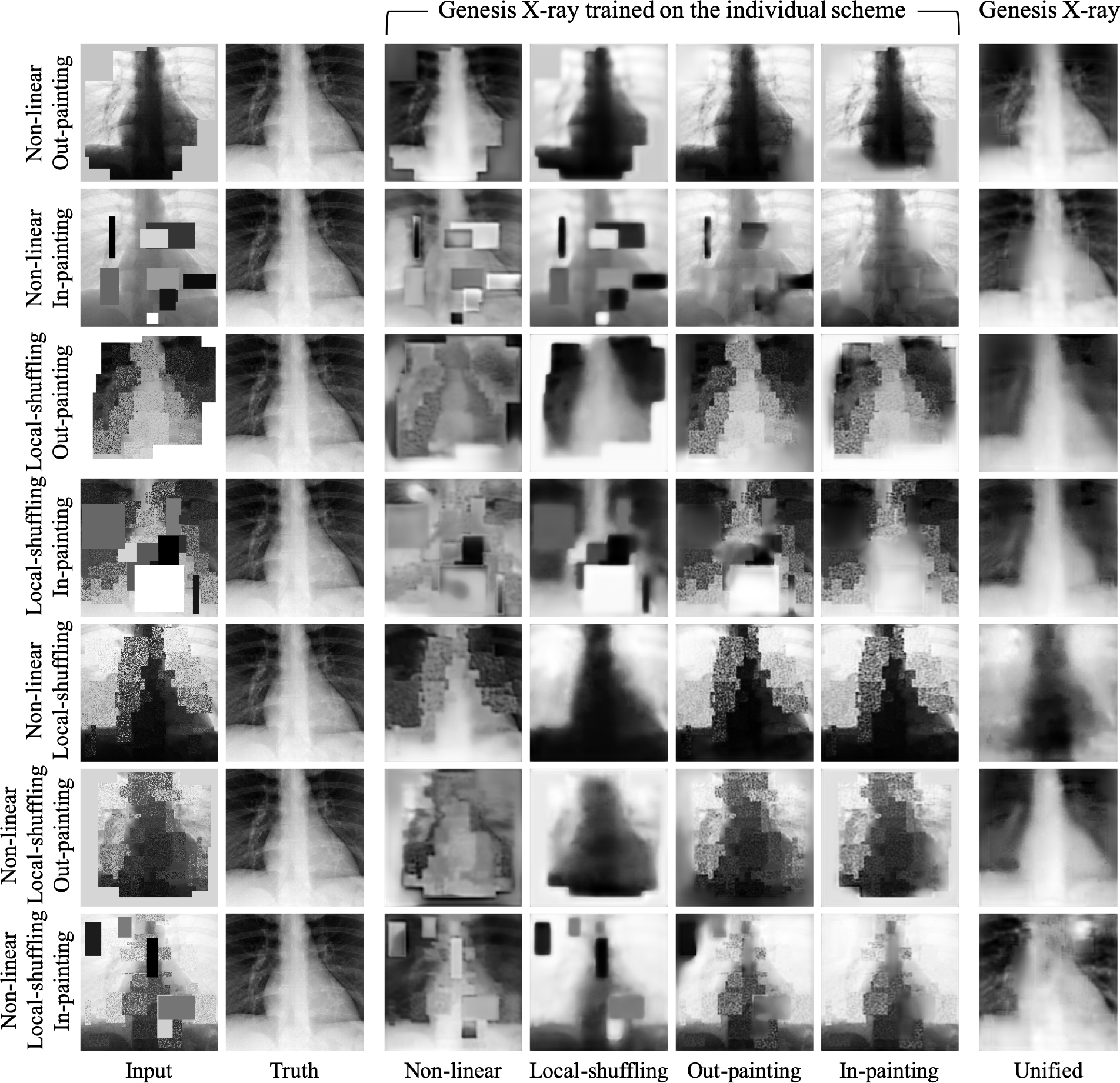}
\caption{
    Continued from~\figurename~\ref{fig:genesis_chest_xray_within_modality_appendix_1}. We further test our five models on seven combinations of transformations. The models train with individual training scheme can only handle a single transformation (Columns 3---6) and fail to restore the patches completely, while Models Genesis trained via proposed unified self-supervised learning framework (Column 7) fairly handle seven augmented transformations and restores the patches close to the original patch. Taking a combination of out-painting and non-linear transformation (Row 1) as an example,  the model trained on non-linear-based scheme (Column 3) recovers the original intensity values, but fails to out-paint the image; however, the model trained with a unified framework not only recovers the original intensity values but also out-paints the image. This observation demonstrates the superiority of Models Genesis trained with unified self-supervised learning framework.
}
\label{fig:genesis_chest_xray_within_modality_appendix_2}
\end{figure}

\newpage
\begin{figure}[!h]
\centering
\includegraphics[width=1.0\linewidth]{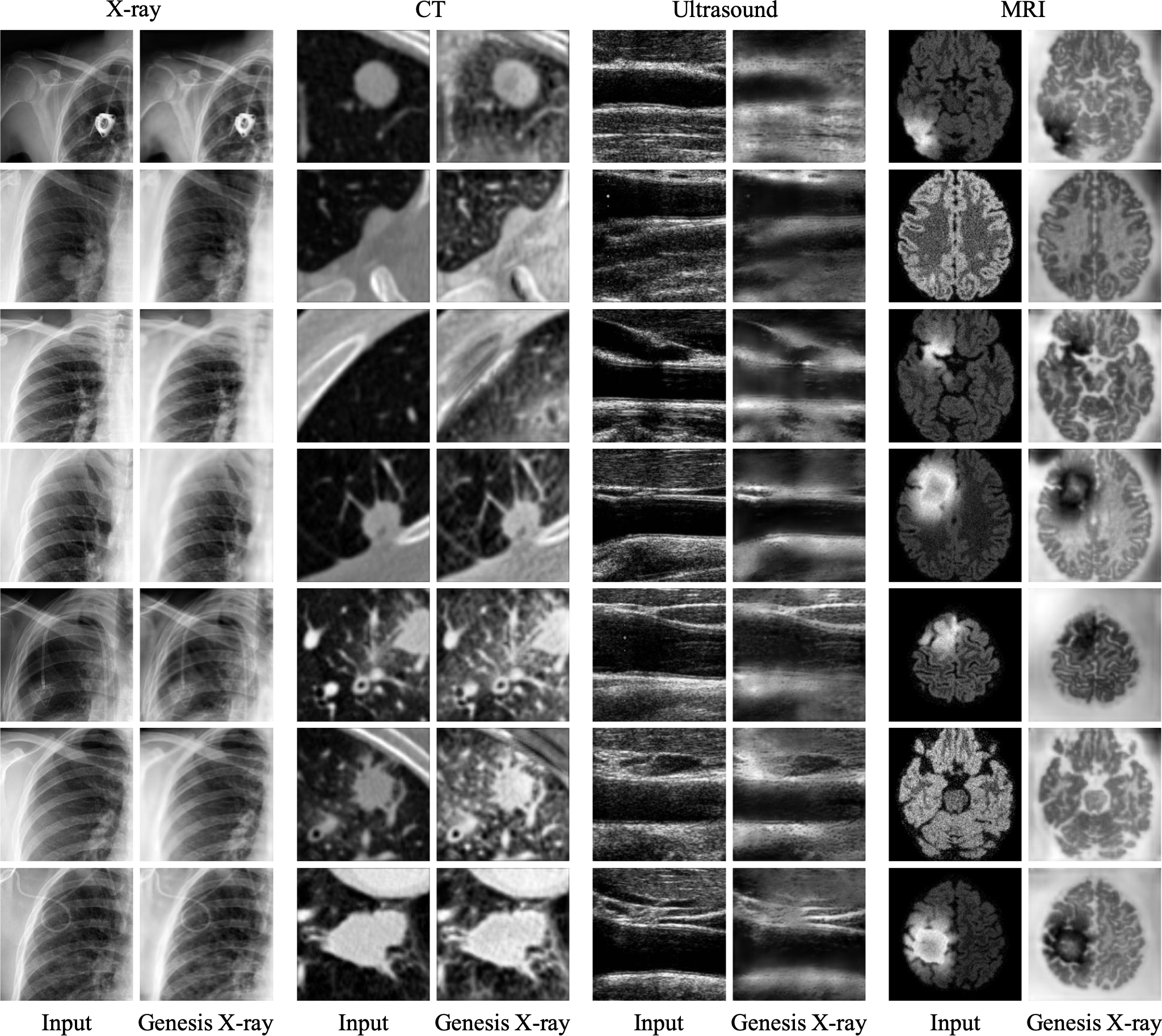}
\caption{
    Qualitative results of image restoration from Genesis Chest X-ray across dataset, organ, and modality are visualized. Genesis Chest X-ray is trained on Chest X-ray8 (X-ray)~\cite{wang2017chestx} via our unified training framework, and tested to restore images from Chest X-ray8 (X-ray), LIDC-IDRI (CT)~\cite{armato2011lung}, CIMT (Ultrasound)~\cite{hurst10,zhou2019integrating}, and BraTS (MRI)~\cite{menze2015multimodal}. Similar to Fig.~\ref{fig:genesis_chest_ct_cross_modality_appendix}, we observe that the performance of restoration and target tasks in various modalities may be positively correlated. For instance, while Genesis Chest X-ray restores ultrasound images  reasonably, it injects unintended artifacts in the restored CT slices. As a result, Genesis Chest X-ray achieves better performance on \texttt{IUC} task compared with Genesis Chest CT, but it fails on \texttt{NCC} task (see Fig.~\ref{fig:2D_target_tasks}).  The analysis further confirms our claims provided in Fig.~\ref{fig:genesis_chest_ct_cross_modality_appendix}.
}
\label{fig:genesis_chest_xray_cross_modality_appendix}
\end{figure}

\newpage
\section{Models Genesis vs. Models ImageNet}
\label{sec:genesis_vs_imagenet_appendix}

\begin{figure}[!h]
\footnotesize
\centering
    \includegraphics[width=0.328\linewidth]{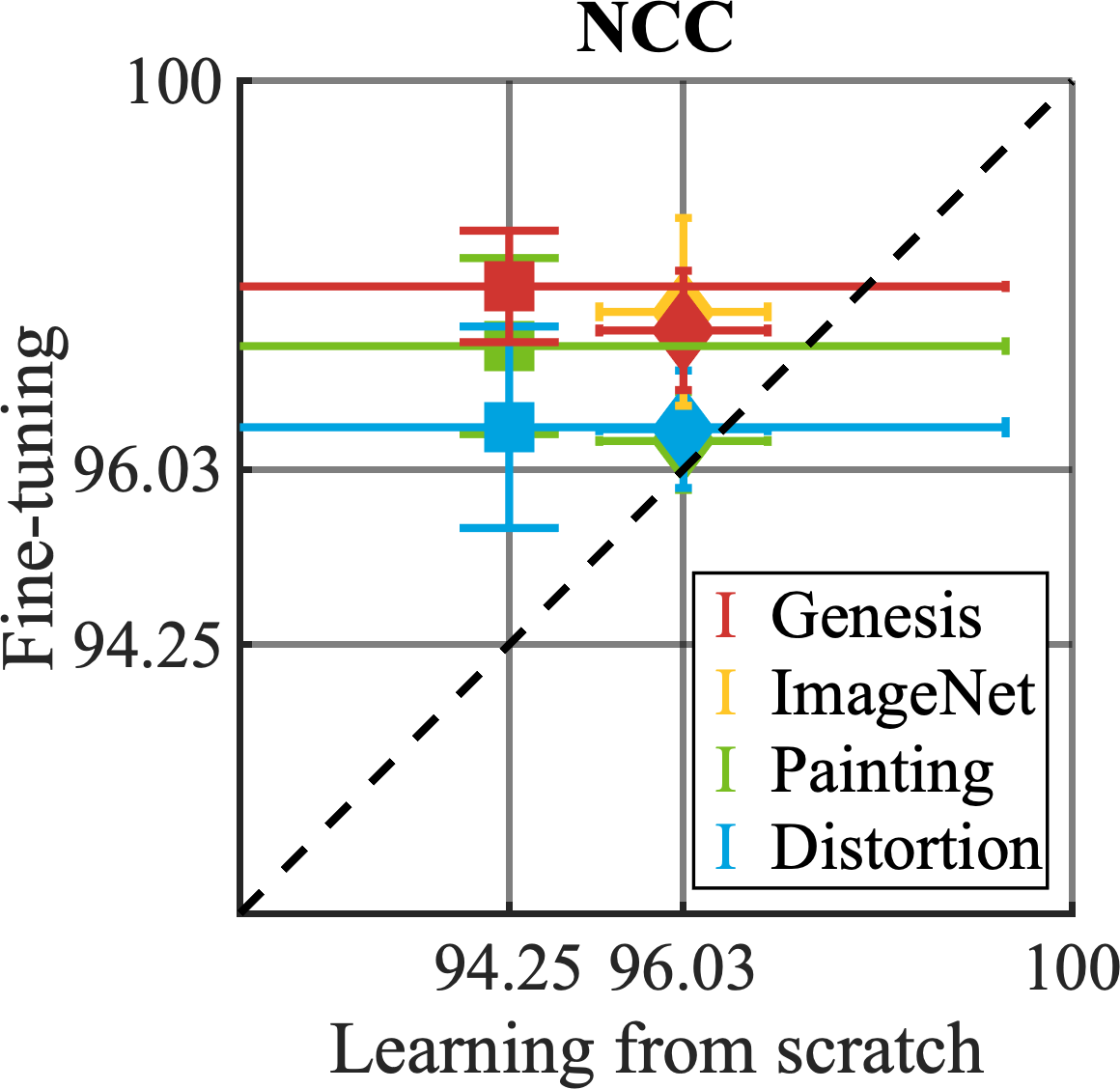}
    \includegraphics[width=0.328\linewidth]{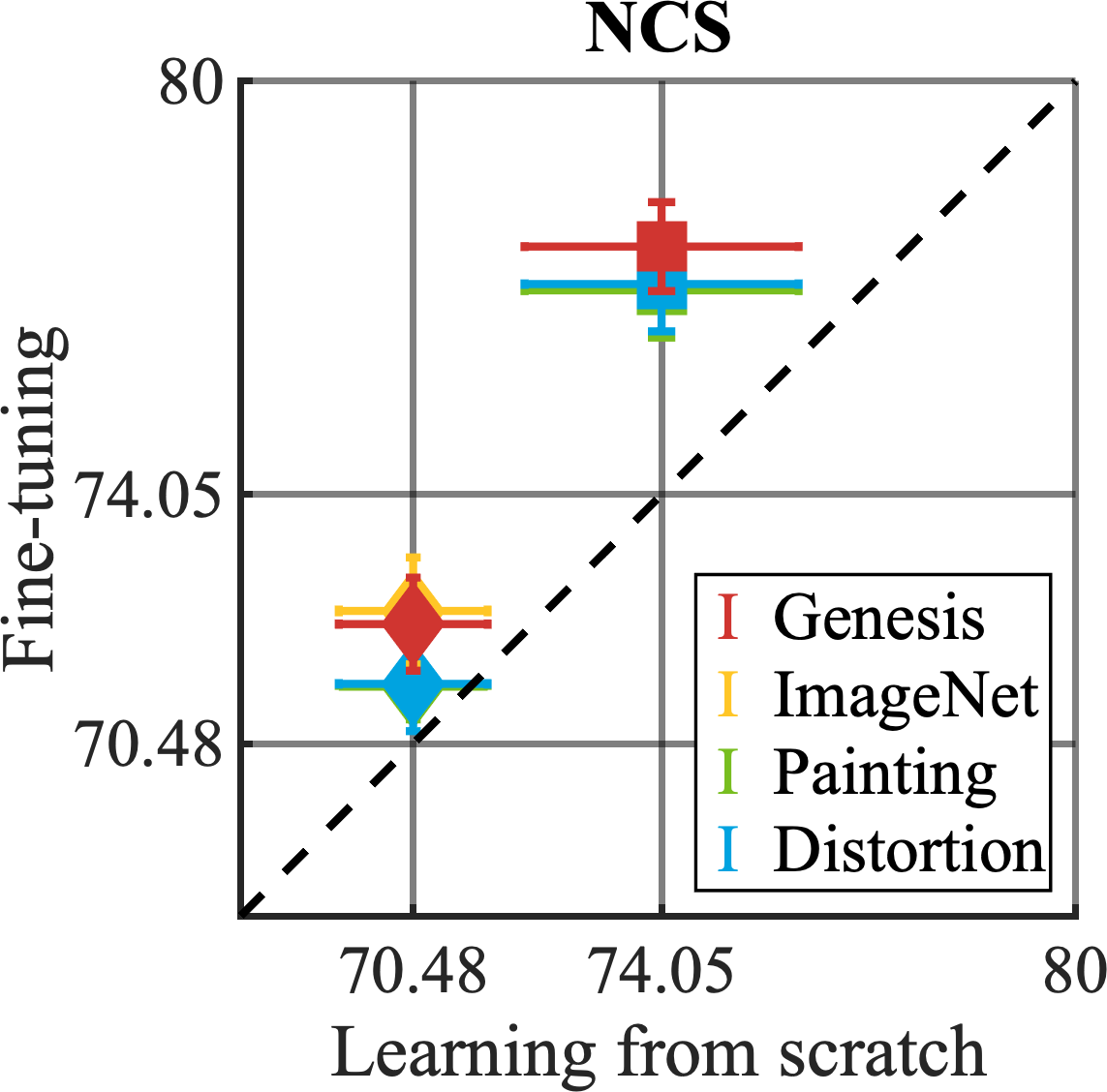}
    \includegraphics[width=0.328\linewidth]{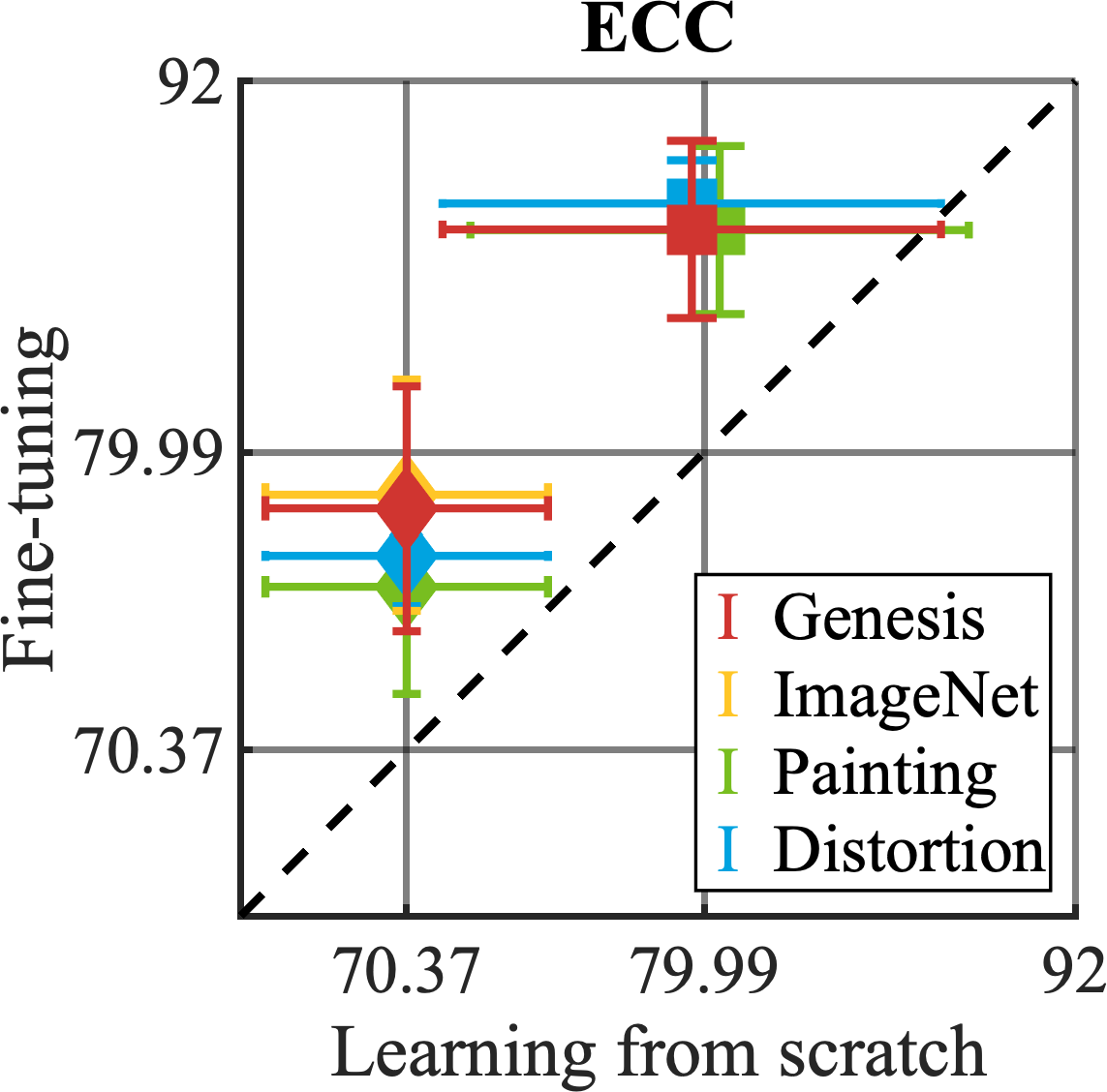}
   
    \scriptsize{
    \begin{tabular}{P{0.07\linewidth}P{0.13\linewidth}P{0.13\linewidth}P{0.13\linewidth}P{0.000001\linewidth}P{0.13\linewidth}P{0.1\linewidth}P{0.13\linewidth}P{0.1\linewidth}}
    \hline
    \multirowcell{2}{Task} & \multicolumn{3}{c}{2D ($\%$)} & & \multicolumn{3}{c}{3D ($\%$)} & \multirowcell{2}{$p$-value$^{\dagger}$} \\
    \cline{2-4}\cline{6-8}
     & Scratch & ImageNet & Genesis & & Scratch & ImageNet & Genesis & \\
    \hline
    \texttt{NCC} & 96.03$\pm$0.86 & \cellcolor{maroon!15} 97.79$\pm$0.71 & 97.45$\pm$0.61 & & 94.25$\pm$5.07 & N/A & \cellcolor{maroon!15}\textbf{98.20$\pm$0.51} & 0.0213 \\
    \texttt{NCS} & 70.48$\pm$1.07 & \cellcolor{maroon!15}72.39$\pm$0.77 & 72.20$\pm$0.67 & & 74.05$\pm$1.97 & N/A & \cellcolor{maroon!15}\textbf{77.62$\pm$0.64} & $<$1$e$-8 \\
    \texttt{ECC} & 71.27$\pm$4.64 & \cellcolor{maroon!15}78.61$\pm$3.73 &  78.58$\pm$3.67 & & 79.99$\pm$8.06 & N/A & \cellcolor{maroon!15}\textbf{88.04$\pm$1.40} & 5.50$e$-4 \\
    \hline
    \end{tabular}
    }
    \begin{tablenotes}
        \scriptsize
        \item $^{\dagger}$These $p$-values are calculated between our Models Genesis vs. the fine-tuning from ImageNet, which always offers the best performance (highlighted in red) for all three tasks in 2D.
    \end{tablenotes}
    
    \scriptsize
    \begin{tabular}{p{0.2\linewidth}P{0.15\linewidth}P{0.15\linewidth}P{0.15\linewidth}P{0.15\linewidth}P{0.15\linewidth}}
    \hline
    Approach & \texttt{NCC} ($\%$) & \texttt{NCS} ($\%$) & \texttt{ECC} ($\%$)& \texttt{LCS} ($\%$)& \texttt{BMS} ($\%$) \\
    \hline
    Scratch & 94.25$\pm$5.07 & 74.05$\pm$1.97 & 79.99$\pm$8.06 & 74.60$\pm$4.57 & 90.16$\pm$0.41 \\
    Distortion (ours) & 96.46$\pm$1.03 & \cellcolor{maroon!15}77.08$\pm$0.68 & \cellcolor{maroon!15}\textbf{88.04$\pm$1.40} & \cellcolor{maroon!15}79.08$\pm$4.26 & \cellcolor{maroon!15}\textbf{90.60$\pm$0.20} \\
    Painting (ours) & \cellcolor{maroon!15}\textbf{98.20$\pm$0.51} & 77.02$\pm$0.58 & 87.18$\pm$2.72 & 78.62$\pm$4.05 & 90.46$\pm$0.21 \\
    Unified (ours) & \cellcolor{maroon!15}97.90$\pm$0.57 & \cellcolor{maroon!15}\textbf{77.62$\pm$0.64} & \cellcolor{maroon!15}87.20$\pm$2.87 & \cellcolor{maroon!15}\textbf{79.52$\pm$4.77} & \cellcolor{maroon!15}90.59$\pm$0.21 \\
    \hline
    $p$-value$^{\dagger\dagger}$ & 0.0848 & 0.0520 & 0.2102 & 0.4249 & 0.4276 \\
    \hline
    \end{tabular}
    \begin{tablenotes}
        \scriptsize
        \item $^{\dagger\dagger}$These $p$-values are calculated between the top-2 models in each column {\zzred highlighted in red}.
    \end{tablenotes}
    
\caption{
    Comparison of Models Genesis and Models ImageNet.
    In the top three sub-figures, the 3D volume-based solutions and 2D slice-based solutions are denoted with square and diamond markers, respectively.
    The horizontal and vertical error bars indicate 95$\%$ confidence intervals of training from scratch and fine-tuning, respectively. The shorter the vertical bar, the more consistent and stable the model is. 
} 

\label{fig:2D_3D_target_tasks_appendix}
\end{figure}

The comparisons of our Models Genesis and Models ImageNet (\ie models pre-trained on ImageNet) are summarized in three figures and two tables in~\figurename~\ref{fig:2D_3D_target_tasks_appendix}.
Training 3D models simply from scratch does not necessarily outperform the 2D counterparts (see \texttt{NCC}), however, fine-tuning the same 3D models from Genesis Chest CT significantly outperforms ($p<0.05$) the slice-based 2D models including fine-tuning from Models ImageNet. As seen, Models Genesis enjoys a higher stability  on the target tasks. Moreover, comparing our unified framework with individual training schemes demonstrates that the former is more robust across all target tasks, yielding either the best result or comparable performance to the best model ($p<0.05$). This superiority of our Models Genesis is attributable to consolidating multiple self-supervised training schemes, which enables the model to learn a stronger image representation.  Thus, fine-tuning Models Genesis leads to powerful and stable application-specific target models, confirming the importance of Models Genesis in 3D medical imaging.

\newpage
\section{The NiftyNet Transfer Learning Capability}
\label{sec:niftynet_appendix}

\begin{figure}[!h]
\scriptsize
\centering
    \begin{tabular}{p{0.35\linewidth}P{0.2\linewidth}P{0.2\linewidth}P{0.2\linewidth}}
    \hline
    Initialization & \texttt{NCS} (Dice $\%$) & \texttt{LCS} (Dice $\%$) & \texttt{BMS} (Dice $\%$) \\
    \hline
    Models Genesis (ours) & 75.86$\pm$0.90 & 91.13$\pm$1.51 & 92.58$\pm$0.30 \\
    \hline
    NiftyNet scratch~\cite{Gibson2018} & 69.65$\pm$2.56 & 91.09$\pm$0.76 & 90.68$\pm$0.24 \\
    NiftyNet model zoo~\cite{gibson2018automatic} & 69.24$\pm$1.77 & 90.84$\pm$0.63 & 90.65$\pm$0.54 \\
    $p$-value$^{\dagger}$ & 0.3433 & 0.2214 & 0.4301 \\
    \hline
    \end{tabular}
    \begin{tablenotes}
        \item {\scriptsize $^{\dagger}$These $p$-values are calculated between NiftyNet scratch and NiftyNet model zoo.}
    \end{tablenotes}
    \includegraphics[width=0.328\linewidth]{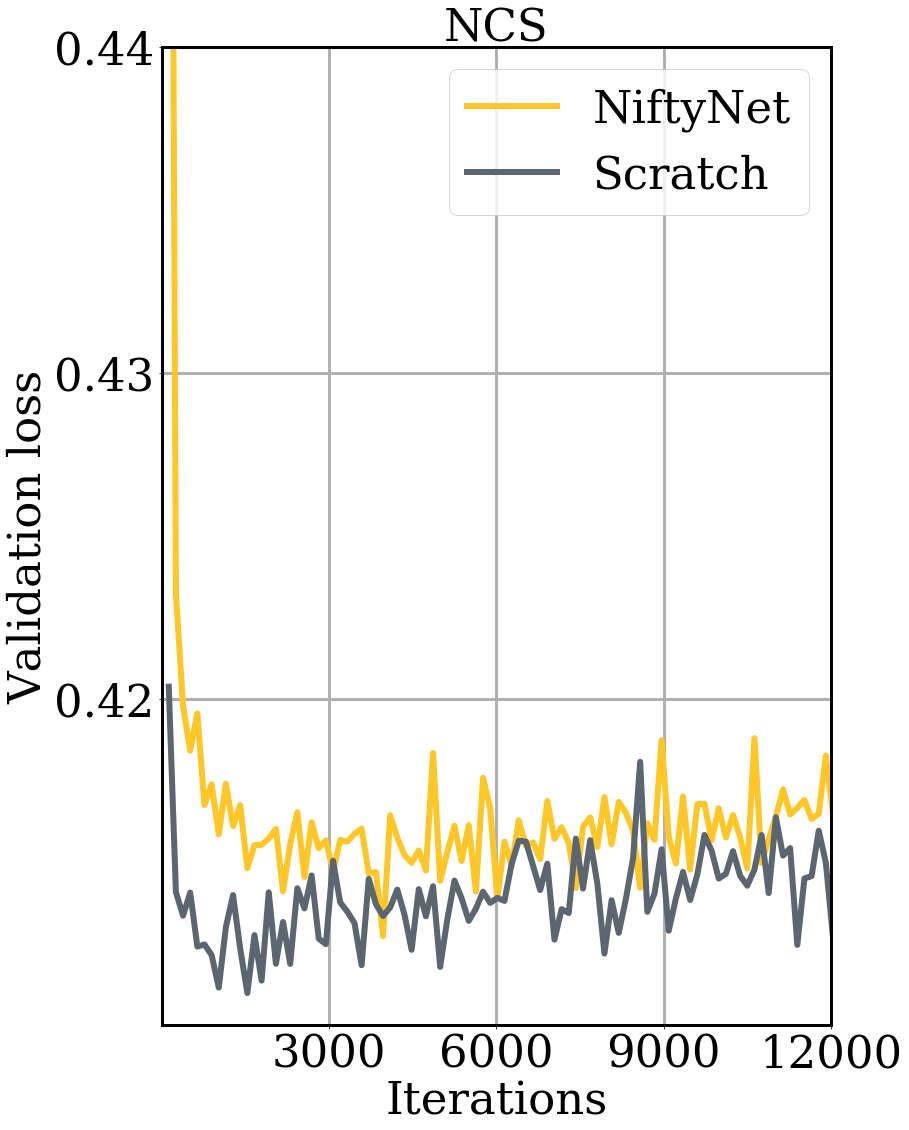}
    \includegraphics[width=0.328\linewidth]{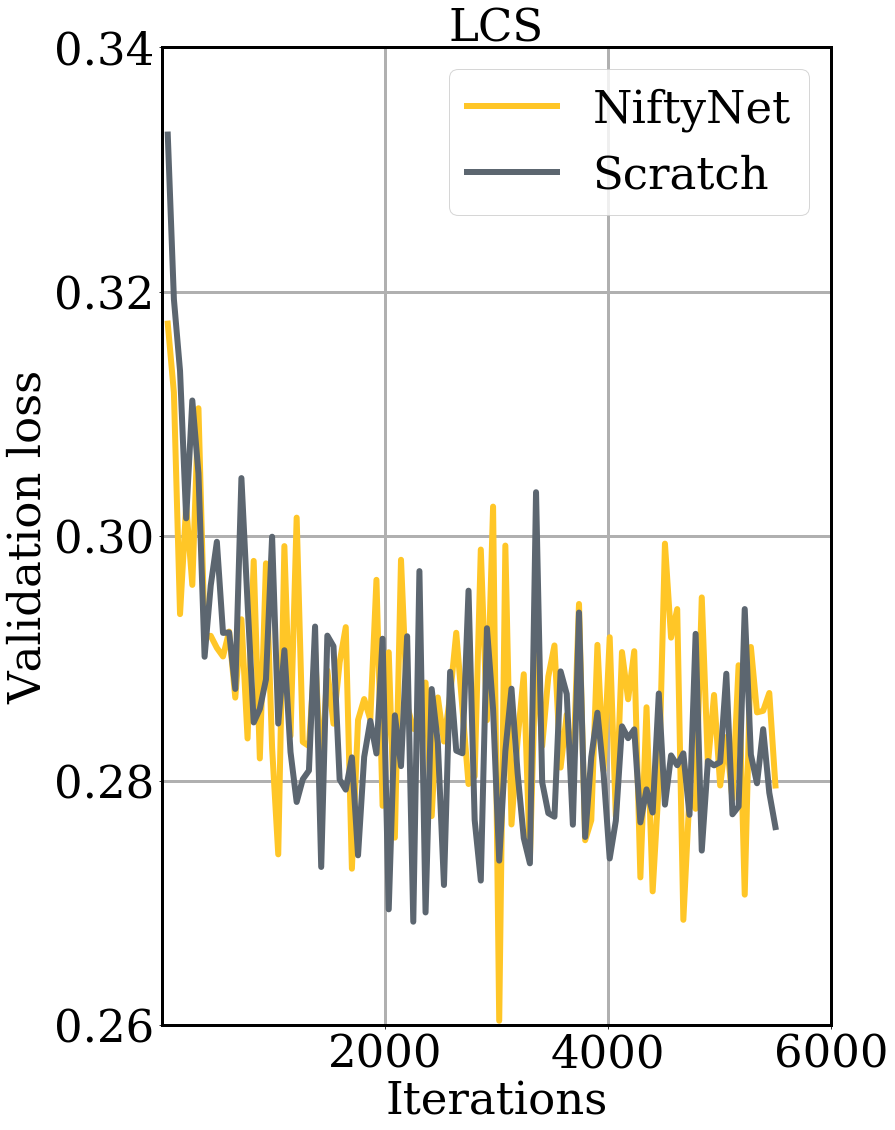}
    \includegraphics[width=0.328\linewidth]{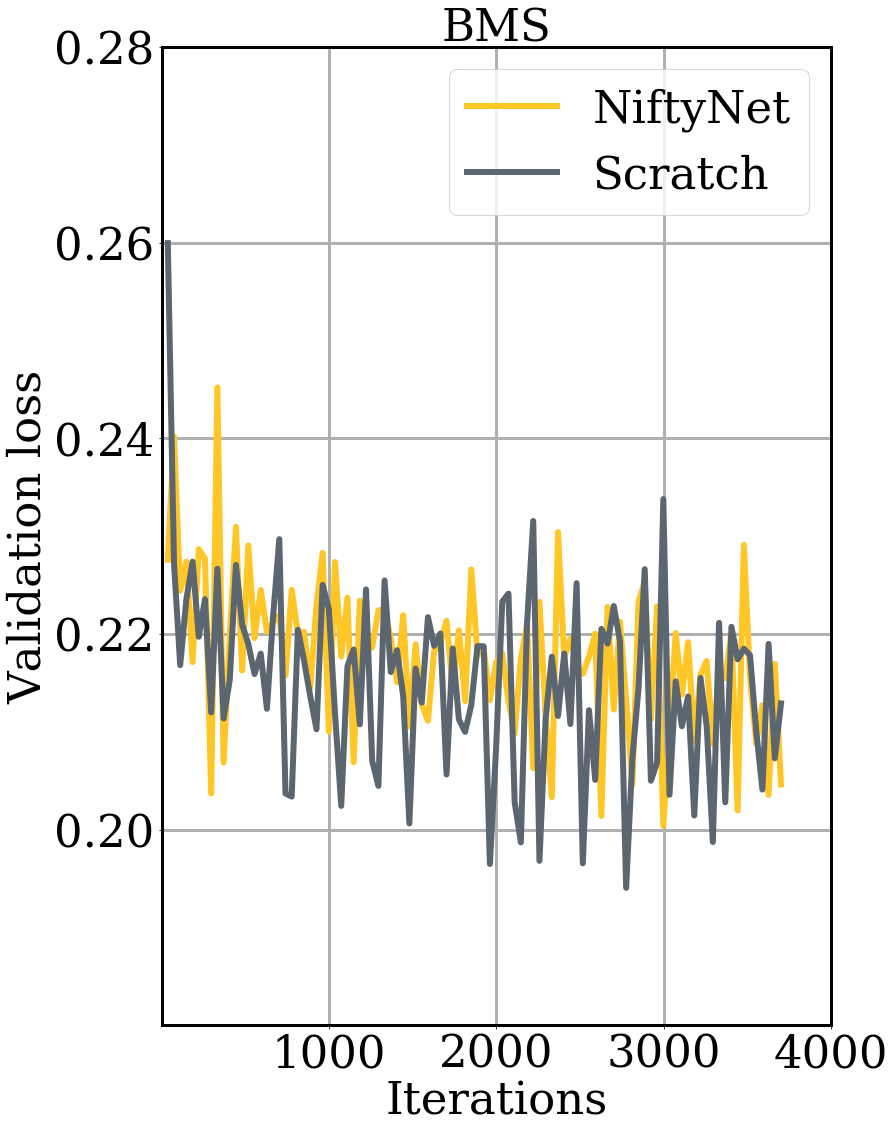}
\caption{
    Fine-tuning the pre-trained NiftyNet vs. training it from scratch. The results reported in the table statistically infer that fine-tuning the pre-trained NiftyNet offers no benefit over training it from scratch. This fact is further supported by the learning curve comparison provided in the figures.
}
\label{fig:niftynet_appendix}
\end{figure}

The table in~\figurename~\ref{fig:niftynet_appendix} compares fine-tuning the pre-trained NiftyNet with training from scratch on three target tasks: (1) lung nodule segmentation (\texttt{NCS}) in CT images, (2) liver segmentation (\texttt{LCS}) in CT images, and (3) brain tumor segmentation (\texttt{BMS}) in MRI images using dice-coefficient (mean$\pm$s.d.) as the evaluation metric, demonstrating  that fine-tuning NiftyNet's 3D supervised pre-trained weights has {\em no} benefit over random initialization ($p > 0.05$). It is further corroborated by the learning curves on validation dataset provided at the bottom in~\figurename~\ref{fig:niftynet_appendix}. However, Models Genesis significantly improve performance over random initialization (see~\tablename~\ref{tab:3D_target_tasks} in the main paper) and perform consistently better than NiftyNet models on the three same target tasks. Note that the pre-trained NiftyNet model was trained using strong supervision, whereas Models Genesis learns representations using the proposed self-supervised paradigm. In contrast to pre-trained weights of NiftyNet's model zoo, the pre-trained weights from our proposed self-supervised method are found to be more robust across diseases, organs, and imaging modalities, thanks to the ability of our approach to learn representations from a large-scale unannotated dataset.



\end{document}